\documentclass[preprint, superscriptaddress, showpacs,preprintnumbers,amsmath,amssymb,prd]{revtex4}
\usepackage{graphicx}
\usepackage{amsmath}\usepackage{slashed}

\begin{document}

\thispagestyle{empty}

\title{Quantum field theoretical description for the reflectivity of graphene }

\author{M.~Bordag}
\affiliation{Institute for Theoretical
Physics, Leipzig University, Postfach 100920,
D-04009, Leipzig, Germany}

\author{
G.~L.~Klimchitskaya}
\affiliation{Central Astronomical Observatory at Pulkovo of the Russian Academy of Sciences, St.Petersburg,
196140, Russia}
\affiliation{Institute of Physics, Nanotechnology and
Telecommunications, St.Petersburg State
Polytechnical University, St.Petersburg, 195251, Russia}

\author{
V.~M.~Mostepanenko}
\affiliation{Central Astronomical Observatory at Pulkovo of the Russian Academy of Sciences, St.Petersburg,
196140, Russia}
\affiliation{Institute of Physics, Nanotechnology and
Telecommunications, St.Petersburg State
Polytechnical University, St.Petersburg, 195251, Russia}

\author{
V.~M.~Petrov}
\affiliation{Institute of Physics, Nanotechnology and
Telecommunications, St.Petersburg State
Polytechnical University, St.Petersburg, 195251, Russia}

\begin{abstract}
We derive the polarization tensor of graphene at nonzero temperature
in (2+1)-dimensional space-time. The obtained tensor coincides with the
 previously known one at all Matsubara frequencies, but, in contrast to
 it, admits  analytic continuation to the real frequency axis
 satisfying all physical requirements. Using the obtained representation
 for the polarization tensor, we develope quantum field theoretical
 description for the reflectivity of graphene. The analytic asymptotic
 expressions for the reflection coefficients and reflectivities at low
 and high frequencies are derived for both independent polarizations of
 the electromagnetic field. The dependencies of reflectivities on the
 frequency and angle of incidence are investigated. Numerical
 computations using the exact expressions for the polarization tensor
 are performed and application
 regions for the analytic asymptotic results are determined.
\end{abstract}

\maketitle

\section{Introduction}

During the last few years, a two-dimensional sheet of carbon atoms called graphene
attracted much experimental and theoretical attention \cite{1}. It is unique material
whose electronic excitations at small frequencies below a few eV demonstrate the linear
dispersion relation and are described by relativistic quantum electrodynamics \cite{1,2}.
This allows calculation of the reflectivity properties of graphene with respect to
incident electromagnetic waves starting from first principles of quantum field theory.
In fact, graphene presents the possibility to test at a laboratory table several effects
predicted by quantum field theory, which were commonly considered unfeasible for
observation in the past due to extreme values of parameters required for their
realization. One could mention the interaction of graphene with an electrostatic
potential barrier (the Klein paradox) \cite{3} and creation of quasiparticle pairs in
graphene either by the Schwinger mechanism in static electric field \cite{4,5} or in
a time-dependent field \cite{6}.

Previously, the reflectivity of transverse magnetic (TM), i.e., $p$ polarized,
electromagnetic waves on graphene was investigated using the local model for the
in-plane (longitudinal) graphene conductivity \cite{7,8}.
Both the transverse magnetic and transverse electric (TE)
 reflection coefficients (i.e., for both, $p$ and $s$,
  polarizations of the electromagnetic
 field) were expressed in terms of the polarization tensor at zero \cite{9} and
 nonzero \cite{10} temperature. This approach, entirely based on quantum
 electrodynamics, was used to calculate the Casimir interaction in different
 systems involving graphene \cite{9,10,11,12,13,14,15,16,17} and turned out to be
 very productive. Specifically, the existence of large thermal effect in the
 Casimir force for graphene at short separations, discovered in Ref.~\cite{18}
 using the longitudinal density-density correlation function in the random phase
 approximation, was confirmed and investigated in detail.

 It should be taken into account, however, that the temperature-dependent
 contribution to the polarization tensor found in Ref.~\cite{10} is perfectly
 correct only at the imaginary Matsubara frequencies.
 The representation of Ref.~\cite{10} uses the Feynman parameter and
 its formal analytic continuation to the
 whole plane of complex frequency, for instance,  to real frequencies,
 disagrees with that used in general thermal
 quantum field theory \cite{19}. Thus, the results of Ref.~\cite{10} are well
 adapted to calculate the Casimir force, but are not applicable for the
 investigation of reflectivity of graphene at real frequencies.
 As a result, calculations of the reflectivity properties of graphene in
 Ref.~\cite{20} using the polarization tensor was limited to the case of
 sufficiently high frequencies, where one can omit the temperature-dependent
 part of this tensor.

 The present paper develops complete quantum field theoretical description
 for the reflectivity of graphene over the entire range of real frequencies
 from zero to infinity. For this purpose we derive another representation for
 the polarization tensor of graphene in (2+1)-dimensional space-time which is
 valid not only at the imaginary Matsubara frequencies, but over the whole
 complex plane, specifically, along the real frequency axis.
  Since the beginning of 1960s it is known which
 representation of the polarization tensor allows for the correct analytic
 continuation \cite{24}. Here we follow methods suggested in
 Refs.~\cite{24,23}.
 In so doing, we compare our representation with that obtained in Ref.~\cite{10}
 and demonstrate quite a different behavior everywhere with exception of only
 the pure imaginary Matsubara frequencies. Furthermore, we consider the scattering
 frequency interval and find analytic
 asymptotic expressions for the polarization tensor and for both  TM and TE
 reflectivities of  graphene. The cases of
 low and high frequencies of incident electromagnetic waves with respect to
 the temperature are investigated.
 We also perform numerical computations of reflectivities and
 find the application regions of our asymptotic expressions. Finally, we compare
 quantum field theoretical results of this paper with those obtained previously \cite{7,8}
 using the local model for the conductivity of graphene.
 A good agreement is demonstrated for the TM reflectivities at both high and
 low frequencies. At the same time, the TE reflectivity is shown to depend on the
 angle of incidence on the contrary to a qualitative conclusion made
 previously \cite{8}.

 The paper is organized as follows. In Sec.~II we derive the polarization tensor
 for graphene applicable over the entire plane of complex frequency.
 Section~III is devoted to the comparison with previously obtained expressions
 for the polarization tensor. The analytic asymptotic expressions for the
 polarization tensor and for both TM and
 TE reflectivities of graphene at high frequencies are obtained in Sec.~IV.
 In Sec.~V the same is done at low frequencies and the comparison with
 the results of numerical
 computations is presented. Section~VI contains our conclusions and discussion.
 The Appendix contains details of derivation of the polarization tensor in
 Sec.~II.

 Throughout the paper we use units in which $\hbar=c=k_B=1$, where $k_B$
is the Boltzmann constant. The fundamental constants are restored for better
understanding where necessary.

\section{The polarization tensor for graphene at nonzero temperature}
\newcommand{\be}{\begin{equation}}\newcommand{\ee}{\end{equation}}
\newcommand{\bea}{\begin{eqnarray}}\newcommand{\eea}{\end{eqnarray}}
\newcommand{\nn}{\nonumber}
\newcommand{\pa}{\partial}
\newcommand{\la}{\lambda}\newcommand{\om}{\omega}
\newcommand{\Ga}{\Gamma}\newcommand{\Gat}{\tilde{\Gamma}}
\newcommand{\ep}{\varepsilon}
\renewcommand{\phi}{\varphi}
\newcommand{\p}{\mathbf{p}}
\newcommand{\q}{\mathbf{q}}
\renewcommand{\k}{\mbox{\boldmath$k$}_{\bot}}
\newcommand{\x}{\mathbf{x}}\newcommand{\xx}{{\rm xx}}
\newcommand{\Ref}[1]{(\ref{#1})}
\renewcommand{\ni}{\noindent}
\renewcommand{\v}{v_{\rm F}}
\newcommand{\elm}{electromagnetic~}
\newcommand{\al}{\alpha}
\newcommand{\etah}{\hat{\eta}}\newcommand{\pt}{\tilde{p}}
\newcommand{\kb}{k_{\bot}}
\newcommand{\qb}{q_{\bot}}

According to the Dirac model, the interaction of the electromagnetic field with
the long wavelength (low frequency) electronic excitations in graphene can be
described by relativistic quantum electrodynamics in (2+1)-dimensions, where
the speed of light inside the graphene sheet is replaced with the Fermi
velocity $\v\approx 1/300$ \cite{1,2}. In this formalism, the Dirac equation is
\begin{equation}
(i  {\slashed \pa}-e\slashed A -m)\psi=0,
\label{eq1}
\end{equation}
\noindent
where $m$ is the mass-gap parameter, $A_{\mu}$ is the vector-potential of the
electromagnetic field,
\begin{equation}
\slashed \pa=\tilde{\gamma}^\mu\frac{\pa}{\pa x^\mu},\quad
\slashed A=\tilde{\gamma}^{\,\mu} A_\mu,
\quad \tilde{\gamma}^{\,\mu}=\eta_{\,\,\nu}^{\,\mu}\gamma^{\,\nu},
\label{eq2}
\end{equation}
\noindent
the Greek indices $\mu,\,\nu=0,\,1,\,2$, $\gamma^{\,\mu}$ are the standard
$\gamma$-matrices and the metric tensor $\eta_{\,\,\nu}^{\mu}$ is given by
\begin{equation}
\eta_{\,\,\nu}^{\,\mu}={\rm diag}(1,\v,\v).
\label{eq3}
\end{equation}
\noindent
Note that we assume graphene situated at the plane $(x^1,x^2)$.

The polarization tensor for graphene at zero temperature was obtained in
Ref.~\cite{9}. In Ref.~\cite{10} it was found  at any nonzero temperature
at all Matsubara frequencies using the Feynman parameter for a massive
fermionic field and earlier, in the same representation, in Ref.~\cite{23a}
for a massless field, but with chemical potential.
Here, we obtain another representation for the polarization tensor of graphene
which is valid over the whole plane of complex frequencies including the real
and imaginary frequency axes.

We start from the definition for the polarization tensor of graphene in the
one-loop approximation in momentum representation \cite{10,21}
\begin{eqnarray}
&&
{\Pi}^{\mu\nu}(k)={\Pi}^{\mu\nu}(k_0,\mbox{\boldmath$k$}_{\bot})
\nonumber \\
&&~~
        =8\pi i\alpha\int\frac{dq_0d\mbox{\boldmath$q$}}{(2\pi)^3}\,
        {\rm tr} \frac{1}{i\slashed q-m}\tilde{\gamma}^{\,\mu} \frac{1}{i\slashed q-i\slashed k-m}\tilde{\gamma}^{\,\nu}.
\label{eq4}
\end{eqnarray}
\noindent
Here, $k\equiv k^{\mu}=(k_0,\mbox{\boldmath$k$}_{\bot})\equiv
(\omega,\mbox{\boldmath$k$}_{\bot})$ is the 3-momentum of an external photon,
$k_0=\omega$ is the frequency, $\mbox{\boldmath$k$}_{\bot}=(k^1,k^2)$,
and $\alpha=e^2\approx 1/137$ in the fine structure constant.
In a similar way, the 3-momentum of a loop electronic excitation is
$q\equiv q^{\mu}=(q_0,\mbox{\boldmath$q$}_{\bot})$,
$\mbox{\boldmath$q$}_{\bot}=(q^1,q^2)$ and we use the notations
$\slashed q=\tilde{\gamma}^{\,\mu}q_{\mu}$,
$\slashed k=\tilde{\gamma}^{\,\mu}k_{\mu}$.
Note that for the purpose of calculating the polarization tensor, the use of
the standard $4\times 4$ $\gamma$-matrices instead of $2\times 2$
$\sigma$-matrices  and an additional multiple 2 included in front of the
right-hand side of Eq.~(\ref{eq4}) take into account $N=4$ number of
fermion species for graphene.

We are interested to find the polarization tensor of graphene at any
temperature $T$. In order to introduce temperature within the Matsubara
formalism, one should replace the integration over $q_0$ in Eq.~(\ref{eq4}) with the
summation over the imaginary fermionic Matsubara frequencies
\begin{equation}
q_{0n}\equiv iq_{4n}=2\pi i\left(n+\frac{1}{2}\right)T,
\label{eq5}
\end{equation}
\noindent
where $n=0,\,\pm 1,\,\pm 2,\,\ldots\,$.
This replacement has the form
\begin{equation}
\int_{-\infty}^{\infty}\frac{dq_0}{2\pi}\to iT\!\sum\limits_{n=-\infty}^{\infty}.
\label{eq6}
\end{equation}
\noindent
One should also replace the frequency $k_0=\omega$ in the argument of
$\Pi^{\mu\nu}$ in Eq.~(\ref{eq4}) with the imaginary bosonic Matsubara
frequencies
\begin{equation}
k_{0l}=\omega_l\equiv ik_{4l}\equiv i\xi_l=2\pi ilT,
\label{eq6a}
\end{equation}
\noindent
where $l$ is an integer number. Later it will be possible to turn the
imaginary frequency back to real frequency.

At both zero and nonzero temperature (as well as at nonzero chemical potential
$\mu$), the polarization tensor is transversal, i.e., it holds
\be
k_\mu \Pi^{\mu\nu}(k_0,\k)=0.
\label{eq7}\ee
\noindent
Another useful property is that in the presence of graphene all components
of the polarization tensor can be expressed in terms of two independent
quantities with corresponding form factors \cite{10}. Following Ref.~\cite{10},
we use the 00-component and the trace of the polarization tensor
\be \Pi_{00}(\omega,\k),\qquad \Pi_{\rm tr}(\omega,\k)=\Pi_{\mu}^{\,\,\mu}(k_0,\k),
\label{eq8}\ee
as independent quantities in further calculations. It is only in the special case $T=0$
(here and below we put $\mu=0$) that the quantities (\ref{eq8}) are connected
by the equation
\be \Pi_{\rm tr}(\omega,\k)=-\frac{2\omega^2-k_{\bot}^2(1+\v^2)}{k_{\bot}^2}\Pi_{00}(\omega,\k),
\label{eq9}\ee
\noindent
where $\kb=|\k|$. As shown below, the components (\ref{eq8}) of the polarization tensor
depend not on $\k$ but on $\kb$.

Using the Maxwell equations describing an electromagnetic field outside the surface of graphene
and the standard matching conditions it is possible to express the reflection coefficients on
graphene for both independent polarizations via the polarization tensor.
The result is \cite{10}:
\begin{eqnarray}
&&
r_{\rm TM}(\omega,k_{\bot})=\frac{1}{1+Q_{\rm TM}^{-1}(\omega,\kb)},
\nonumber \\
&&
r_{\rm TE}(\omega,k_{\bot})=-\frac{1}{1+Q_{\rm TE}^{-1}(\omega,\kb)},
\label{eq10}
\end{eqnarray}
\noindent
where the quantities $Q_{\rm TM(TE)}$ are given by
\begin{eqnarray}
&&
Q_{\rm TM}(\omega,\kb)=\frac{p(\omega,\kb)}{2i\kb^2}\,\Pi_{00}(\omega,\kb),
\label{eq11} \\
&&
Q_{\rm TE}(\omega,\kb)=-\frac{1}{2ip(\omega,\kb)}\,\Pi_{\rm tr}(\omega,\kb)
-\frac{p(\omega,\kb)}{2i\kb^2}\,\Pi_{00}(\omega,\kb).
\nonumber
\end{eqnarray}
\noindent
Here, we have introduced the notation
\begin{equation}
p(\omega,\kb)=\sqrt{\omega^2-\kb^2}.
\label{eq12}
\end{equation}
\noindent
Substituting Eq.~(\ref{eq11}) in Eq.~(\ref{eq10}), one obtains the reflection
coefficients in terms of the polarization tensor \cite{10},
\begin{eqnarray}
&&
r_{\rm TM}(\omega,k_{\bot})=\frac{p(\omega,\kb)\Pi_{00}}{2i\kb^2+p(\omega,\kb)\Pi_{00}},
\label{eq13} \\[2mm]
&&
r_{\rm TE}(\omega,k_{\bot})=-
\frac{\kb^2\Pi_{\rm tr}+p^2(\omega,\kb)\Pi_{00}}{-2i\kb^2p(\omega,\kb)
+\kb^2\Pi_{\rm tr}+p^2(\omega,\kb)\Pi_{00}},
\nonumber
\end{eqnarray}
\noindent
where, for the sake of brevity, we put
$\Pi_{00({\rm tr})}\equiv\Pi_{00({\rm tr})}(\omega,\kb)$.

Equations (\ref{eq11}) and (\ref{eq13}) define the reflection coefficients (\ref{eq10})
in the Minkowskian region, i.e., for real $\omega$ and $\omega\geq\kb$ such that the
quantity $p(\omega,\kb)$ defined in Eq.~(\ref{eq12})
(it has an interpretation of the wave vector component perpendicular to the
graphene sheet) is also real.
The transition to the Euclidean region is done by putting $\omega=i\xi$.
Then Eq.~(\ref{eq11}) turns into
\begin{eqnarray}
&&
Q_{\rm TM}(i\xi,\kb)=\frac{q(\xi,\kb)}{2\kb^2}\,\Pi_{00}(i\xi,\kb),
\label{eq14} \\
&&
Q_{\rm TE}(i\xi,\kb)=\frac{1}{2q(\xi,\kb)}\,\Pi_{\rm tr}(i\xi,\kb)
-\frac{q(\xi,\kb)}{2\kb^2}\,\Pi_{00}(i\xi,\kb),
\nonumber
\end{eqnarray}
\noindent
where
\begin{equation}
q(\xi,\kb)=\sqrt{\xi^2+\kb^2}=-ip(i\xi,\kb).
\label{eq15}
\end{equation}
\noindent
Substituting Eq.~(\ref{eq14}) in Eq.~(\ref{eq10}) taken at
$\omega=i\xi$, one obtains
\begin{eqnarray}
&&
r_{\rm TM}(i\xi,k_{\bot})=\frac{q(\xi,\kb)\Pi_{00}}{2\kb^2+q(\xi,\kb)\Pi_{00}},
\label{eq16} \\[2mm]
&&
r_{\rm TE}(i\xi,k_{\bot})=-
\frac{\kb^2\Pi_{\rm tr}-q^2(\xi,\kb)\Pi_{00}}{2q(\xi,\kb)\kb^2
+\kb^2\Pi_{\rm tr}-q^2(\xi,\kb)\Pi_{00}},
\nonumber
\end{eqnarray}
\noindent
where
$\Pi_{00({\rm tr})}\equiv\Pi_{00({\rm tr})}(i\xi,\kb)$.

We are coming now to the calculation of polarization tensor of graphene at
any temperature $T$. It is convenient to represent the independent quantities
(\ref{eq8}) considered at the imaginary Matsubara frequencies in the form
\begin{equation}
\Pi_{00({\rm tr})}(i\xi_l,\k)=\Pi_{00({\rm tr})}^{(0)}(i\xi_l,\k)+
\Delta_T\Pi_{00({\rm tr})}(i\xi_l,\k),
\label{eq17}
\end{equation}
\noindent
where $\Pi_{00({\rm tr})}^{(0)}(i\xi,\k)$ is the 00-component (trace) of the
polarization tensor at $T=0$ but calculated at the imaginary bosonic Matsubara
frequencies $ik_{4l}=i\xi_l$, and $\Delta_T\Pi_{00({\rm tr})}(i\xi_l,\kb)$ is the thermal
correction which explicitly depends on the temperature and goes to zero in
the limiting case $T\to 0$. The explicit expressions for $\Pi_{00({\rm tr})}^{(0)}$
are well known \cite{9}
\begin{eqnarray}
&&
\Pi_{00}^{(0)}(i\xi_l,\kb)=\frac{\alpha\kb^2}{\tilde{q}^{\,2}(\xi_l,\kb)}\,
\Phi(\xi_l,\kb),
\label{eq18} \\
&&
\Pi_{{\rm tr}}^{(0)}(i\xi_l,\kb)=\alpha
\frac{q^2(\xi_l,\kb)+\tilde{q}^{\,2}(\xi_l,\kb)}{\tilde{q}^{\,2}(\xi_l,\kb)}\,
\Phi(\xi_l,\kb),
\nonumber
\end{eqnarray}
\noindent
where
\begin{eqnarray}
&&
\tilde{q}(\xi_l,\kb)=\sqrt{\xi_l^2+\v^2\kb^2},
\label{eq19} \\
&&
\Phi(\xi_l,\kb)=4\left[m+
\frac{\tilde{q}^{\,2}(\xi_l,\kb)-4m^2}{2\tilde{q}(\xi_l,\kb)}\,
{\rm arctan}\frac{\tilde{q}(\xi_l,\kb)}{2m}\right].
\nonumber
\end{eqnarray}
\noindent
The expressions for $\Delta_T\Pi_{00({\rm tr})}$ valid at the imaginary
Matsubara frequencies were obtained in
Ref.~\cite{10}. Here, we derive the representation for $\Delta_T\Pi_{00({\rm tr})}$
which admits the analytic continuation to the real frequency axis (see
Appendix A).
According to Eqs.~(\ref{A.18})--(\ref{A.21}) the result is
\begin{eqnarray}
&&
\Delta_T\Pi_{00({\rm tr})}(i\xi_l,\kb)=\frac{16\alpha}{\v^2}
\int_0^{\infty} d\qb\frac{\qb}{\Gamma(\qb)}\,\frac{1}{e^{\Gamma(\qb)/T}+1}
\nonumber \\
&&~~~~~
\times \left[1+\frac{1}{2}\sum\limits_{\lambda=\pm 1}
\frac{M_{00({\rm tr})}(\xi_l,\kb,\qb)}{N(\xi_l,\kb,\qb)}\right],
\label{eq20}
\end{eqnarray}
\noindent
where $\qb=|\mbox{\boldmath$q$}_{\bot}|$ and the following notations are
introduced:
\begin{eqnarray}
&&
N(\xi_l,\kb,\qb)=\left[Q^2(\xi_l,\kb,\qb)
-(2\v\kb\qb)^2\right]^{1/2},
\nonumber \\
&&
Q(\xi_l,\kb,\qb)=\tilde{q}^{\,2}(\xi_l,\kb)
-2i\lambda\xi_l\Gamma(\qb),
\nonumber \\
&&
\Gamma(\qb)=\sqrt{\qb^2+m^2},
\label{eq21} \\
&&
M_{00}(\xi_l,\kb,\qb)=4\Gamma^2(\qb)-\tilde{q}^{\,2}(\xi_l,\kb)+
4i\lambda\xi_l\Gamma(\qb),
\nonumber \\
&&
M_{\rm tr}(\xi_l,\kb,\qb)=-\tilde{q}^{\,2}(\xi_l,\kb)+4\v^2m^2
\nonumber \\
&&~~~~~~~~~~
+4(1-\v^2)\left[\Gamma^2(\qb)+i\lambda\xi_l\Gamma(\qb)\right].
\nonumber
\end{eqnarray}
\noindent
Note that for all $\xi>0$ including the Matsubara frequencies $\xi_l$ with
$l\geq 1$ the summation in $\lambda=\pm 1$ in Eq.~(\ref{eq20}) can be equivalently
represented as
\begin{equation}
\frac{1}{2}\sum\limits_{\lambda=\pm 1}
\frac{M_{00({\rm tr})}}{N}={\rm Re}\frac{M_{00({\rm tr})}}{N}.
\label{eq22}
\end{equation}
\noindent
We emphasize that in order to obtain the correct result in Eq.~(\ref{eq20}) at
$\xi_0=0$, one should first make the summation in $\lambda=\pm 1$  at $\xi>0$
and then consider the limit $\xi\to 0$. Note that if one puts $\xi_0=0$ first,
the subsequent summation in $\lambda=\pm 1$ would lead to a complex result for
some relationships between $\kb$ and $\qb$ in contradiction with the fact that
the polarization tensor along the imaginary frequency axis must be real.
The proposed continuation of the expression (\ref{eq20}) to zero Matsubara
frequency is equivalent to the use of Eq.~(\ref{eq22}) at all $\xi\geq 0$.
It is also confirmed in Sec.~III by the comparison with another representation
for the polarization tensor of graphene.

In the massless case, $m=0$, the above representation for  the polarization tensor of graphene
can be simplified. Using Eqs.~(\ref{eq18}) and (\ref{eq19}), for the zero temperature
parts one obtains
\begin{eqnarray}
&&
\Pi_{00}^{(0)}(i\xi_l,\kb)=\frac{\pi\alpha\kb^2}{\tilde{q}(\xi_l,\kb)},
\label{eq23} \\
&&
\Pi_{{\rm tr}}^{(0)}(i\xi_l,\kb)=\pi\alpha
\frac{q^2(\xi_l,\kb)+\tilde{q}^{\,2}(\xi_l,\kb)}{\tilde{q}(\xi_l,\kb)}.
\nonumber
\end{eqnarray}
\noindent
For the thermal correction in the massless case $\Gamma(\qb)=\qb$ and from
Eq.~(\ref{eq21}) we have
\begin{equation}
N(\xi_l,\kb,\qb)=\tilde{q}(\xi_l,\kb)W(\xi_l,\kb,\qb),
\label{eq24}
\end{equation}
\noindent
where
\begin{equation}
W(\xi_l,\kb,\qb)=\left[\tilde{q}^{\,2}(\xi_l,\kb)-4i\lambda\xi_l\qb
-4\qb^2\right]^{1/2}.
\label{eq24a}
\end{equation}
\noindent
In a similar way, from Eq.~(\ref{eq21}) in the massless case one arrives at
\begin{eqnarray}
&&
M_{00}(\xi_l,\kb,\qb)=-W^2(\xi_l,\kb,\qb),
\label{eq25} \\
&&
M_{\rm tr}(\xi_l,\kb,\qb)=-(1-\v^2)W^2(\xi_l,\kb,\qb)
-\v^2\tilde{q}^{\,2}(\xi_l,\kb).
\nonumber
\end{eqnarray}
\noindent
Substituting Eqs.~(\ref{eq24}) and (\ref{eq25}) in Eq.~(\ref{eq20}), we find
the thermal correction to the polarization tensor in the massless case
\begin{eqnarray}
&&
\Delta_T\Pi_{00}(i\xi_l,\kb)=\frac{16\alpha}{\v^2}
\int_0^{\infty} d\qb\frac{1}{e^{{\qb}/T}+1}
\nonumber \\
&&~~~~~
\times \left[1-\frac{1}{2}\sum\limits_{\lambda=\pm 1}
\frac{W(\xi_l,\kb,\qb)}{\tilde{q}(\xi_l,\kb)}\right],
\label{eq26} \\
&&
\Delta_T\Pi_{\rm tr}(i\xi_l,\kb)=\frac{16\alpha}{\v^2}
\int_0^{\infty} d\qb\frac{1}{e^{{\qb}/T}+1}
\nonumber \\
&&~~~~
\times\left\{1-\frac{1}{2}\sum\limits_{\lambda=\pm 1}\left[
(1-\v^2)\frac{W(\xi_l,\kb,\qb)}{\tilde{q}(\xi_l,\kb)}
+\v^2\frac{\tilde{q}(\xi_l,\kb)}{W(\xi_l,\kb,\qb)}\right]
\right\} .
\nonumber
\end{eqnarray}

Finally, we consider the Minkowskian region by turning the imaginary frequency to the
real frequency axis in accordance to $\omega=i\xi$. Here we have several frequency
subintervals. For the lowest frequencies, $\omega\leq\v\kb$, all the above formulas
hold literally and $\tilde{q}$ remains real after putting $\xi^2=-\omega^2$.
The interval where surface plasmons may exist goes next. It is characterized by
\begin{eqnarray}
&&
\v\kb<\omega<\kb,
\label{eq27}\\
&&
q(\omega,\kb)=\sqrt{-\omega^2+\kb^2},\qquad
\tilde{p}(\omega,\kb)=\sqrt{\omega^2-\v^2\kb^2},
\nonumber
\end{eqnarray}
\noindent
where $q$ and $\tilde{p}$ are real. The scattering interval
\begin{equation}
\kb\leq\omega
\label{eq28}
\end{equation}
\noindent
is the most important one because it corresponds to real photons on a
mass-shell. Here, both the quantity
$p(\omega,\kb)$ defined in Eq.~(\ref{eq12}) and
$\tilde{p}(\omega,\kb)$ defined in Eq.~(\ref{eq27}) are real.

In the plasmonic and in the scattering frequency intervals, the
polarization tensor has a similar form. At zero temperature it
is given by
\begin{eqnarray}
&&
\Pi_{00}^{(0)}(\omega,\kb)=-\frac{\alpha\kb^2}{\tilde{p}^{\,2}(\omega,\kb)}
\,\Phi(\omega,\kb),
\label{eq29} \\
&&
\Pi_{{\rm tr}}^{(0)}(\omega,\kb)=\alpha
\frac{p^2(\omega,\kb)+
\tilde{p}^{\,2}(\omega,\kb)}{\tilde{p}^{\,2}(\omega,\kb)}\,\Phi(\omega,\kb),
\nonumber
\end{eqnarray}
\noindent
where the function $\Phi$ along the real frequency axis is defined as
\begin{equation}
\Phi(\omega,\kb)=\left\{
\begin{array}{ll}
4\left[m-
\frac{\tilde{p}^{\,2}(\omega,\kb)+4m^2}{2\tilde{p}(\omega,\kb)}\,
{\rm arctanh}\frac{\tilde{p}(\omega,\kb)}{2m}\right],\quad &
\omega\leq\sqrt{\v^2\kb^2+4m^2}, \\[2mm]
4\left[m-
\frac{\tilde{p}^{\,2}(\omega,\kb)+4m^2}{2\tilde{p}(\omega,\kb)}\,
\left({\rm arctanh}\frac{2m}{\tilde{p}(\omega,\kb)}
+\frac{i\pi}{2}\right)\right],\quad &
\omega>\sqrt{\v^2\kb^2+4m^2}.
\end{array}
\right.
\label{eq30}
\end{equation}
\noindent
The thermal correction to the polarization tensor in the plasmonic
and scattering frequency intervals takes the form of Eq.~(\ref{eq20}),
\begin{eqnarray}
&&
\Delta_T\Pi_{00({\rm tr})}(\omega,\kb)=\frac{16\alpha}{\v^2}
\int_0^{\infty} d\qb\frac{\qb}{\Gamma(\qb)}\,\frac{1}{e^{\Gamma(\qb)/T}+1}
\nonumber \\
&&~~~~~
\times \left[1+\frac{1}{2}\sum\limits_{\lambda=\pm 1}
\frac{M_{00({\rm tr})}(\omega,\kb,\qb)}{N(\omega,\kb,\qb)}\right].
\label{eq31}
\end{eqnarray}
\noindent
Here, the denominator $N$ takes different forms depending on the value
of the quantity [compare with Eq.~(\ref{eq21})]
\begin{equation}
Q(\omega,\kb,\qb)=-\tilde{p}^{\,2}(\omega,\kb)-2\lambda\omega\Gamma(\qb).
\label{eq32}
\end{equation}
\noindent

If  $|Q(\omega,\kb,\qb)|\geq2\v\kb\qb$ we have
\begin{eqnarray}
&&
N(\omega,\kb,\qb)={\rm sign}Q(\omega,\kb,\qb)
\nonumber\\
&&~~~~~~\times
\left[
Q^2(\omega,\kb,\qb)-(2\v\kb\qb)^2\right]^{1/2},
\label{eq33}
\end{eqnarray}
\noindent
whereas if $|Q(\omega,\kb,\qb)|<2\v\kb\qb$
\begin{equation}
N(\omega,\kb,\qb)=-i
\left[
-Q^2(\omega,\kb,\qb)+(2\v\kb\qb)^2\right]^{1/2}
\label{eq34}
\end{equation}
\noindent
holds. Two other quantities entering Eq.~(\ref{eq31}) are given by
\begin{eqnarray}
&&
M_{00}(\omega,\kb,\qb)=4\Gamma^2(\qb)+\tilde{p}^{\,2}(\omega,\kb)+
4\lambda\omega\Gamma(\qb),
\nonumber \\
&&
M_{\rm tr}(\omega,\kb,\qb)=\tilde{p}^{\,2}(\omega,\kb)+4\v^2m^2
\label{eq35} \\
&&~~~~~~~~
+4(1-\v^2)\left[\Gamma^2(\qb)+\lambda\omega\Gamma(\qb)\right].
\nonumber
\end{eqnarray}
\noindent
Note that in the plasmonic frequency interval the denominator $N$ has
no zeroes, but in the scattering interval with $\lambda=-1$ it has two
zeros at
\begin{equation}
(\qb)_{1,2}=\mp\frac{\v\kb}{2}+\frac{\omega}{2}
\sqrt{1-\frac{4m^2}{\tilde{p}^{\,2}(\omega,\kb)}}.
\label{eq36}
\end{equation}
\noindent
These zeros result in the integrable singularities.

Now we consider the massless case in the scattering interval where
Eqs.~(\ref{eq31}) and (\ref{eq33})--(\ref{eq35}) for the polarization
tensor can be significantly simplified.
At zero temperature the result follows from Eqs.~(\ref{eq29})
and (\ref{eq30}) in the limiting case $m\to 0$ \cite{20},
\begin{eqnarray}
&&
\Pi_{00}^{(0)}(\omega,\kb)=\frac{i\pi\alpha\kb^2}{\tilde{p}(\omega,\kb)},
\label{eq37} \\
&&
\Pi_{{\rm tr}}^{(0)}(\omega,\kb)=-i\pi\alpha
\frac{p^2(\omega,\kb)+\tilde{p}^{\,2}(\omega,\kb)}{\tilde{p}(\omega,\kb)}.
\nonumber
\end{eqnarray}

For the thermal correction in the massless case from Eqs.~(\ref{eq33})
and (\ref{eq34}) one obtains
\begin{equation}
N(\omega,\kb,\qb)=\tilde{p}(\omega,\kb)W(\omega,\kb,\qb),
\label{eq38}
\end{equation}
\noindent
where
\begin{equation}
W^2(\omega,\kb,\qb)=\tilde{p}^{\,2}(\omega,\kb)+4\lambda\omega\qb
+4\qb^2.
\label{eq39}
\end{equation}
\noindent
The explicit expression for $W(\omega,\kb,\qb)$ follows from  Eqs.~(\ref{eq33})
and (\ref{eq34}). It is different for different $\lambda=\pm 1$ and depends
on the value of $\qb$. Thus, for $\lambda=1$ and for any $\qb$ it holds
\begin{equation}
W(\omega,\kb,\qb)=-\left[\tilde{p}^{\,2}(\omega,\kb)+4\omega\qb
+4\qb^2\right]^{1/2},
\label{eq40}
\end{equation}
\noindent
whereas for $\lambda=-1$ we have
\begin{equation}
W(\omega,\kb,\qb)=\left\{
\begin{array}{lc}
-\left[\tilde{p}^{\,2}(\omega,\kb)-4\omega\qb
+4\qb^2\right]^{1/2},\quad &
2\qb\leq \omega-\v\kb, \\[1.5mm]
-i\,\left[-\tilde{p}^{\,2}(\omega,\kb)+4\omega\qb
-4\qb^2\right]^{1/2},\quad &
\omega-\v\kb<2\qb<\omega+\v\kb, \\[1.5mm]
\left[\tilde{p}^{\,2}(\omega,\kb)-4\omega\qb
+4\qb^2\right]^{1/2},\quad &
2\qb\geq\omega+\v\kb.
\end{array}
\right.
\label{eq41}
\end{equation}
\noindent
In so doing all expressions under the square roots in  Eqs.~(\ref{eq40})
and (\ref{eq41}) are positive.

Using Eq.~(\ref{eq39}), the quantities $M_{00}$ and $M_{\rm tr}$ defined in
Eq.~(\ref{eq35}) can be written in the massless case as
\begin{eqnarray}
&&
M_{00}(\omega,\kb,\qb)=W^2(\omega,\kb,\qb),
\label{eq42} \\
&&
M_{\rm tr}(\omega,\kb,\qb)=(1-\v^2)W^2(\omega,\kb,\qb)+
\v^2\tilde{p}^{\,2}(\omega,\kb).
\nonumber
\end{eqnarray}
\noindent
Substituting Eqs.~(\ref{eq38})
and (\ref{eq42}) in Eq.~(\ref{eq31}), we arrive at the thermal correction to
the polarization tensor of graphene with $m=0$ in the scattering interval,
\begin{eqnarray}
&&
\Delta_T\Pi_{00}(\omega,\kb)=\frac{16\alpha}{\v^2}
\int_0^{\infty} d\qb\frac{1}{e^{{\qb}/T}+1}
\nonumber \\
&&~~~~~
\times \left[1+\frac{1}{2}\sum\limits_{\lambda=\pm 1}
\frac{W(\omega,\kb,\qb)}{\tilde{p}(\omega,\kb)}\right],
\label{eq43} \\
&&
\Delta_T\Pi_{\rm tr}(\omega,\kb)=\frac{16\alpha}{\v^2}
\int_0^{\infty} d\qb\frac{1}{e^{{\qb}/T}+1}
\nonumber \\
&&~~~~
\times\left\{1+\frac{1}{2}\sum\limits_{\lambda=\pm 1}\left[
(1-\v^2)\frac{W(\omega,\kb,\qb)}{\tilde{p}(\omega,\kb)}
+\v^2\frac{\tilde{p}(\omega,\kb)}{W(\omega,\kb,\qb)}\right]
\right\} .
\nonumber
\end{eqnarray}
\noindent
Note also that in the scattering interval (\ref{eq38}) under
consideration here, the denominator of $\Delta_T\Pi_{\rm tr}$ has
zeros at
\begin{equation}
(\qb)_{1,2}=\frac{\omega\pm \v\kb}{2}.
\label{eq44}
\end{equation}
\noindent
These singularities are integrable.

\section{Comparison between different representations for the
polarization tensor of graphene}

In previous section, we have obtained the polarization tensor of graphene
at the imaginary Matsubara frequencies $i\xi_l$ and found its analytic
continuation to the whole plane of complex frequencies including the
real frequency axis. As discussed in Sec.~I, another representation for the
polarization tensor is contained in Ref.~\cite{10}. Here we compare both
representations and demonstrate that the tensor of Ref.~\cite{10} can be
used solely at the Matsubara frequencies, but not along the real frequency
axis [i.e., Eqs.~(13) and (24) in Ref.~\cite{10}].
For the sake of brevity, here we restrict ourselves to the case of
gapless graphene ($m=0$), which is used in our calculations of reflectivities
in the following sections.

First of all we note that the polarization tensor of graphene at zero
temperature calculated at the imaginary Matsubara frequencies is given by
Eq.~(\ref{eq23}) in both representations. Because of this, the subjects of
our comparison are only the thermal corrections $\Delta_T\Pi_{00({\rm tr})}$.
Taking into account the structure of the reflection coefficients (\ref{eq16}),
below we compare different representations not for $\Delta_T\Pi_{00}(i\xi_l,\kb)$
and $\Delta_T\Pi_{{\rm tr}}(i\xi_l,\kb)$ but for $\Delta_T\Pi_{00}(i\xi_l,\kb)$
and the following quantity:
\begin{equation}
\Delta_T\Pi(i\xi_l,\kb)\equiv\kb^2\Delta_T\Pi_{\rm tr}(i\xi_l,\kb)-
q^2(\xi_l,\kb)\Delta_T\Pi_{00}(i\xi_l,\kb).
\label{eq45}
\end{equation}
\noindent
In our representation $\Delta_T\Pi_{00}(i\xi_l,\kb)$ is given by the first line of
Eq.~(\ref{eq26}) and the explicit expression for $\Delta_T\Pi(i\xi_l,\kb)$
follows from Eq.~(\ref{eq26}):
\begin{eqnarray}
&&
\Delta_T\Pi(i\xi_l,\kb)=\frac{16\alpha}{\v^2}\int_{0}^{\infty}
\frac{d\qb}{e^{\qb/T}+1}\,\left[
\vphantom{\frac{W(\xi_l,\kb,\qb)}{\tilde{q}(\xi_l,\kb)}}
-\xi_l^2
\right.
\label{eq46} \\
&&~~~~~~\left.
+
\tilde{q}^{\,2}(\xi_l,\kb){\rm Re}
\frac{W(\xi_l,\kb,\qb)}{\tilde{q}(\xi_l,\kb)}
-\v^2\kb^2{\rm Re}
\frac{\tilde{q}(\xi_l,\kb)}{W(\xi_l,\kb,\qb)}\right].
\nonumber
\end{eqnarray}
\noindent
We also take into account Eq.~(\ref{eq22}). Our representation
 should be compared with the following alternative results of
Ref.~\cite{10} (see also Ref.~\cite{14}) which we notate by an upper
index $(a)$:
\begin{eqnarray}
&&
\Delta_T\Pi_{00}^{(a)}(i\xi_l,\kb)=\frac{8\alpha}{\v^2}
\int_{0}^{1}dx\left\{
\vphantom{\frac{\cos(\xi_lx/T)+
e^{-\Theta_T(x,\xi_l,\kb)}}{\cosh\Theta_T(x,\xi_l,\kb)+\cos(\xi_lx/T)}}
T\ln\left[1+2\cos(\xi_lx/T)e^{-\Theta_T(x,\xi_l,\kb)}
\right.\right.
\nonumber \\
&&~~~
\left.
+e^{-2\Theta_T(x,\xi_l,\kb)}\right]-\frac{\xi_l}{2}(1-2x)
\frac{\sin(\xi_lx/T)}{\cosh\Theta_T(x,\xi_l,\kb)+\cos(\xi_lx/T)}
\nonumber \\
&&~~
\left.
+\frac{\xi_l^2\sqrt{x(1-x)}}{\tilde{q}(\xi_l,\kb)}
\frac{\cos(\xi_lx/T)+
e^{-\Theta_T(x,\xi_l,\kb)}}{\cosh\Theta_T(x,\xi_l,\kb)+\cos(\xi_lx/T)}
\right\},
\label{eq47} \\
&&
\Delta_T\Pi^{(a)}(i\xi_l,\kb)=-\frac{8\alpha}{\v^2}
\int_{0}^{1}dx\left\{
\vphantom{\frac{\cos(\xi_lx/T)+
e^{-\Theta_T(x,\xi_l,\kb)}}{\cosh\Theta_T(x,\xi_l,\kb)+\cos(\xi_lx/T)}}
T\xi_l^2\ln\left[1+2\cos(\xi_lx/T)e^{-\Theta_T(x,\xi_l,\kb)}
\right.\right.
\nonumber \\
&&~~~
\left.
+e^{-2\Theta_T(x,\xi_l,\kb)}\right]-
[2\tilde{q}^{\,2}(\xi_l,\kb)-\xi_l^2]\frac{\xi_l}{2}(1-2x)
\frac{\sin(\xi_lx/T)}{\cosh\Theta_T(x,\xi_l,\kb)+\cos(\xi_lx/T)}
\nonumber \\
&&~~
\left.
+\sqrt{x(1-x)}{\tilde{q}^{\,3}(\xi_l,\kb)}
\frac{\cos(\xi_lx/T)+
e^{-\Theta_T(x,\xi_l,\kb)}}{\cosh\Theta_T(x,\xi_l,\kb)+\cos(\xi_lx/T)}
\right\},
\nonumber
\end{eqnarray}
\noindent
where
\begin{equation}
\Theta_T(x,\xi_l,\kb)=\frac{\tilde{q}(\xi_l,\kb)}{T}\sqrt{x(1-x)}.
\label{eq48}
\end{equation}

As shown in Refs.~\cite{10,11,12}, for graphene the major contribution
to the thermal correction comes from the zero Matsubara frequency.
We start from this case and by putting $l=0$, $\xi_0=0$ in Eqs.~(\ref{eq26})
and (\ref{eq46}) obtain
\begin{eqnarray}
&&
\Delta_T\Pi_{00}(0,\kb)=\frac{16\alpha}{\v^2}\left(
\int_{0}^{\infty}\frac{d\qb}{e^{\qb/T}+1}-\frac{1}{\v\kb}
\int_{0}^{\v\kb/2}\!d\qb
\frac{\sqrt{\v^2\kb^2-4\qb^2}}{e^{\qb/T}+1}\right),
\label{eq49}\\
&&
\Delta_T\Pi(0,\kb)={16\alpha}
\int_{0}^{\v\kb/2}\frac{d\qb}{e^{\qb/T}+1}\left(
-\frac{\v\kb^3}{\sqrt{\v^2\kb^2-4\qb^2}}+
\frac{\kb}{\v}\sqrt{\v^2\kb^2-4\qb^2}\right).
\nonumber
\end{eqnarray}
\noindent
By introducing the new variable $z=2\qb/(\v\kb)$, these results can be
rewritten in the form
\begin{eqnarray}
&&
\Delta_T\Pi_{00}(0,\kb)=\frac{16\alpha}{\v^2}\left(
T\ln 2-\frac{\v\kb}{2}
\int_{0}^{1}\!dz
\frac{\sqrt{1-z^2}}{e^{Az/2}+1}\right),
\nonumber\\
&&
\Delta_T\Pi(0,\kb)={-8\alpha}\v\kb^3
\int_{0}^{1}\frac{dz}{e^{Az/2}+1}
\frac{z^2}{\sqrt{1-z^2}},
\label{eq50}
\end{eqnarray}
\noindent
where $A\equiv\v\kb/T$.

The alternative expressions of Ref.~\cite{10} are obtained
by putting $l=0$, $\xi_0=0$ in Eq.~(\ref{eq47})
\begin{eqnarray}
&&
\Delta_T\Pi_{00}^{(a)}(0,\kb)=\frac{8\alpha T}{\v^2}
\int_{0}^{1}\!dx
\ln\left[1+2e^{-A\sqrt{x(1-x)}}+e^{-2A\sqrt{x(1-x)}}\right],
\label{eq51}\\
&&
\Delta_T\Pi^{(a)}(0,\kb)={-8\alpha}\v\kb^3
\int_{0}^{1}dx\sqrt{x(1-x)}
\frac{1+e^{-A\sqrt{x(1-x)}}}{\cosh[A\sqrt{x(1-x)}]+1}.
\nonumber
\end{eqnarray}

By making transformations in the first line of Eq.~(\ref{eq51}) and
integrating by parts, we have
\begin{eqnarray}
&&
\Delta_T\Pi_{00}^{(a)}(0,\kb)=\frac{16\alpha T}{\v^2}
\int_{0}^{1}\!dx
\ln\left[1+e^{-A\sqrt{x(1-x)}}\right]
\nonumber \\
&&~~
=\frac{16\alpha T}{\v^2}\left[T\ln 2+\v\kb
\int_{0}^{1/2}dx\frac{x(1-2x)}{2\sqrt{x(1-x)}}\,\frac{1}{e^{A\sqrt{x(1-x)}}+1}
\right.
\nonumber \\
&&~~~\left.
+\v\kb
\int_{1/2}^{1}dx\frac{x(1-2x)}{2\sqrt{x(1-x)}}\,\frac{1}{e^{A\sqrt{x(1-x)}}+1}
\right].
\label{eq52}
\end{eqnarray}
\noindent
In both these integrals we introduce the new variable
\begin{eqnarray}
&&
y=2\sqrt{x(1-x)},\qquad dy=\frac{1-2x}{\sqrt{x(1-x)}},
\nonumber \\
&&
x_{1,2}=\frac{1\pm\sqrt{1-y^2}}{2},
\label{eq53}
\end{eqnarray}
\noindent
where the signs plus and minus correspond to the second and first integrals,
respectively. This leads to the result coinciding with the first line of
Eq.~(\ref{eq50}).

In a similar way, the second line of Eq.~(\ref{eq51}) can be represented in the form
\begin{eqnarray}
&&
\Delta_T\Pi^{(a)}(0,\kb)=-{16\alpha}{\v\kb^3}
\int_{0}^{1}\!dx
\frac{\sqrt{x(1-x)}}{e^{A\sqrt{x(1-x)}}+1}
\nonumber \\
&&~~
=-{16\alpha}{\v\kb^3}\left[
\int_{0}^{1/2}dx\frac{\sqrt{x(1-x)}}{e^{A\sqrt{x(1-x)}}+1}
+\int_{1/2}^{1}dx\frac{\sqrt{x(1-x)}}{e^{A\sqrt{x(1-x)}}+1}
\right].
\label{eq54}
\end{eqnarray}
\noindent
After introducing the new variable (\ref{eq53}), where
the sign plus corresponds to the second integral and the sign minus to the first,
Eq.~(\ref{eq54}) coincides with the second line in
Eq.~(\ref{eq50}). Thus, it holds
\begin{equation}
\Delta_T\Pi_{00}(0,\kb)=\Delta_T\Pi_{00}^{(a)}(0,\kb)
\qquad
\Delta_T\Pi(0,\kb)=\Delta_T\Pi^{(a)}(0,\kb).
\label{eq55}
\end{equation}

In fact our representation for the polarization tensor in Eqs.~(\ref{eq26})
and (\ref{eq46}) coincides with that of Ref.~\cite{10} [see Eq.~(\ref{eq47})]
not only at zeroth but at all Matsubara frequencies.
To illustrate this, in the second column of Table~I we present the values
of $\Delta_T\Pi_{00}^{(a)}$, normalized by the coefficient
$C\equiv 16\alpha T/\v^2$, which are computed using the first formula of
Eq.~(\ref{eq47}) at $\kb=10\xi_1$ and first eleven nonzero Matsubara
frequencies indicated in the first column.
In the third column of this table, our results for $\Delta_T\Pi_{00}$
computed using the first formula of Eq.~(\ref{eq26}) are presented.
As can be seen in Table~I, both sets of results coincide up to a high
accuracy. The fourth column contains the relative thermal
correction $\Delta_T\Pi_{00}/\Pi_{00}^{(0)}$, which decreases quickly
with increasing $l$. Similar computations were performed at
$\kb=0.5\xi_1$, $100\xi_1$ and $500\xi_1$. It was found that in all
cases the values of $\Delta_T\Pi_{00}(i\xi_l,\kb)$ computed using
different representations coincide. Note that with respect to the
Casimir effect \cite{25,26} the value $\kb=10\xi_1$ is the magnitude
of the wave vector giving the major contribution to the force between
two parallel graphene sheets spaced at the separation distance 100\,nm.

In Table~II similar computational results for the quantities
$\Delta_T\Pi(i\xi_l,\kb)$ and $\Delta_T\Pi^{(a)}(i\xi_l,\kb)$
are presented normalized by the coefficient
$D\equiv 8\alpha\xi_1^3/\v^2$.
As can be seen in this table, the thermal correction $\Delta_T\Pi$
computed using the representation of Ref.~\cite{10} [i.e., the second
formula of Eq.~(\ref{eq47}), column 2] and the approach of this paper,
i.e., Eq.~(\ref{eq46}) (column 3) coincide at all Matsubara
frequencies (column 1). The same values of $\kb$, as in Table~I,
were used in computations. The relative thermal correction is given
in column 4.
Thus, one can conclude that
\begin{equation}
\Delta_T\Pi_{00}(i\xi_l,\kb)=\Delta_T\Pi_{00}^{(a)}(i\xi_l,\kb)
\qquad
\Delta_T\Pi(i\xi_l,\kb)=\Delta_T\Pi^{(a)}(i\xi_l,\kb).
\label{eq56}
\end{equation}

It can be easily seen, however, that the polarization tensors of graphene
obtained in Ref.~\cite{10} and in this paper behave quite differently at all
frequencies other than the Matsubara frequencies. To see this, in Fig.~1 we
plot $10^5\times\Delta_T\Pi_{00}/C$ (a) and $\Delta_T\Pi_{00}^{(a)}/C$ (b)
as functions of $\xi/\xi_1$ at $\kb=10\xi_1$. In so doing, the quantities
$\Delta_T\Pi_{00}$ and $\Delta_T\Pi_{00}^{(a)}$ were computed using the first
formulas of Eqs.~(\ref{eq26}) and (\ref{eq47}), respectively, where the
Matsubara frequencies $\xi_l$ are replaced with the continuous frequency $\xi$
[see Eqs.~(32) and (33) of Ref.~\cite{27} where the polarization tensor of
Ref.~\cite{10} is written along continuous pure imaginary frequencies].
As is seen in Fig.~1(a,b), our thermal correction decreases monotonously with
increasing $\xi$, whereas the thermal correction of Ref.~\cite{10} oscillates
between the values it takes at the Matsubara frequencies.
This is a nonphysical analytic continuation from the set of Matsubara
frequencies to the imaginary frequency axis and, as a consequence, to the
whole plane of complex frequencies including the real frequency axis
(see Sec.~IV for the physical results).

In Fig.~2(a,b) we present the respective results for the quantity
$\Delta_T\Pi(i\xi,\kb)$ at $\kb=10\xi_1$. Computations were performed
in the framework of our formalism by Eq.~(\ref{eq46}) [Fig.~2(a)] and using
the formalism of Ref.~\cite{10} by the second formula of Eq.~(\ref{eq47})
[Fig.~2(b)]. The quantities $10^4\times \Delta_T\Pi/D$ and
$\Delta_T\Pi^{(a)}/D$ are again plotted as functions of continuous
frequency along the imaginary frequency axis. In our formalism,
dependence of the thermal correction $\Delta_T\Pi$ on the frequency is
again monotonous, whereas the quantity $\Delta_T\Pi^{(a)}$ oscillates
taking the same values as $\Delta_T\Pi$ only at the Matsubara frequencies.
The respective continuation of $\Delta_T\Pi^{(a)}$ to the real frequency
axis does not possess necessary physical properties (see Sec.~IV).
This is an example of the well known general situation in the complex
analysis that a sequence $f(l)$ with no accumulation point  can give rise
to different analytic functions $F(z)$ such that $F(l)=f(l)$.
For example, the function $\Gamma(z+1)$ provides an analytic continuation
of $n!$ to the entire complex plane, but this continuation is not unique.
In particular, the analytic function, e. g., of the form
$\Gamma(z+1)+C\sin(\pi z)$, where $C=\rm const$, also provides the
desired continuation.

\section{Reflectivity of graphene at high frequencies}

Here, we find analytic asymptotic expressions for the polarization tensor
of graphene in the scattering frequency interval (\ref{eq28}) and
respective graphene reflectivities at high frequencies satisfying the
condition $\omega\gg T$. If we restore the fundamental constants,
this condition takes the form
\begin{equation}
\omega\gg\omega_T\equiv\frac{k_BT}{\hbar},
\label{eq57}
\end{equation}
\noindent
where $\omega_T$ is so called thermal frequency. For example, at room
temperature $T=300\,$K the thermal frequency
$\omega_T\approx 3.9\times 10^{13}\,$rad/s, i.e., Eq.~(\ref{eq57}) is well
applicable in the optical range and at all higher frequencies.

Calculations are done starting from Eq.~(\ref{eq43}) for
$\Delta_T\Pi_{00}(\omega,\kb)$ along the real frequency axis and the following
expression for $\Delta_T\Pi(\omega,\kb)$, entering the reflection coefficient
$r_{\rm TE}$ in Eq.~(\ref{eq13}), which is obtained from Eq.~(\ref{eq43}):
\begin{eqnarray}
&&
\Delta_T\Pi(\omega,\kb)=\kb^2\Delta_T\Pi_{\rm tr}(\omega,\kb)+
p^2(\omega,\kb)\Delta_T\Pi_{00}(\omega,\kb)
\nonumber \\
&&~
=\frac{16\alpha}{\v^2}\int_{0}^{\infty}
\frac{d\qb}{e^{\qb/T}+1}\,\left[
\vphantom{\frac{W(\xi_l,\kb,\qb)}{\tilde{q}(\xi_l,\kb)}}
\omega^2
\right.
\label{eq58} \\
&&~~\left.
+
\frac{\tilde{p}^{\,2}(\omega,\kb)}{2}\sum\limits_{\lambda=\pm 1}
\frac{W(\omega,\kb,\qb)}{\tilde{p}(\omega,\kb)}
+\frac{\v^2\kb^2}{2}\sum\limits_{\lambda=\pm 1}
\frac{\tilde{p}(\omega,\kb)}{W(\omega,\kb,\qb)}\right].
\nonumber
\end{eqnarray}
\noindent
At zero temperature the quantity $\Pi_{00}^{(0)}$ is given by the
first line in Eq.~(\ref{eq37}) and the quantity $\Pi^{(0)}$ is given by
\begin{eqnarray}
&&
\Pi^{(0)}(\omega,\kb)=\kb^2\Pi_{\rm tr}^{(0)}(\omega,\kb)+p^2(\omega,\kb)
\Pi_{00}^{(0)}(\omega,\kb)
\nonumber \\
&&~~~~~~~~~
=-i\pi\alpha\kb^2\tilde{p}(\omega,\kb).
\label{eq59}
\end{eqnarray}
\noindent
We note that at $\kb=0$ the quantity $\Delta_T\Pi_{00}(\omega,0)=0$.
This is seen from Eqs.~(\ref{eq40}), (\ref{eq41}) and (\ref{eq43})
if to take into account that  at $\lambda=1$
$W(\omega,0,\qb)=-\omega-2\qb$ whereas at $\lambda=-1$
$W(\omega,0,\qb)=2\qb-\omega$ holds for any $\qb$. Then, by using Eq.~(\ref{eq58}),
one concludes that  $\Delta_T\Pi(\omega,0)=0$ as well. As can be seen from
Eqs.~(\ref{eq37}) and (\ref{eq59}), the same property
$\Pi_{00}^{(0)}(\omega,0)= \Pi^{(0)}(\omega,0)=0$ holds at zero temperature
as well.

Now we obtain the asymptotic expression for $\Delta_T\Pi_{00}$ under the
condition (\ref{eq57}). The main contribution to the integral in the first
formula of Eq.~(\ref{eq43}) is given by $\qb\sim T$. Taking into account
that $\omega\gg T$ and $\omega\geq \kb$, one can conclude that the second and
third intervals of $\qb$ in Eq.~(\ref{eq41}) do not contribute to the result
essentially. Keeping in mind that for $\qb\sim T$
\begin{equation}
\omega-2\qb\approx\omega\gg\v\kb
\label{eq60}
\end{equation}
\noindent
holds, for $\lambda=1$ from Eq.~(\ref{eq40}) we obtain
\begin{equation}
W_1\equiv W(\omega,\kb,\qb)\approx -(\omega+2\qb)+
\frac{\v^2\kb^2}{2(\omega+2\qb)}.
\label{eq61}
\end{equation}
\noindent
Similarly, from the first line of Eq.~(\ref{eq41}) we find
\begin{equation}
W_{-1}\equiv W(\omega,\kb,\qb)\approx -(\omega-2\qb)+
\frac{\v^2\kb^2}{2(\omega-2\qb)}.
\label{eq62}
\end{equation}
\noindent
Taking into account that
\begin{equation}
\frac{1}{\tilde{p}(\omega,\kb)}\approx\frac{1}{\omega}\,
\left(1+\frac{\v^2\kb^2}{2\omega^2}\right),
\label{eq63}
\end{equation}
\noindent
from Eqs.~(\ref{eq61})--(\ref{eq63}) one arrives at
\begin{equation}
\frac{1}{2}\,\frac{W_1+W_{-1}}{\tilde{p}(\omega,\kb)}\approx
-1+\frac{2\v^2\kb^2\qb^2}{\omega^4}.
\label{eq64}
\end{equation}
\noindent
Substituting Eq.~(\ref{eq64}) in the first formula of Eq.~(\ref{eq43}),
one obtains the final result for the real part of $\Delta_T\Pi_{00}$:
\begin{eqnarray}
&&
{\rm Re}\Delta_T\Pi_{00}(\omega,\kb)\approx32\alpha\frac{\kb^2}{\omega^4}
\int_{0}^{\infty}d\qb\frac{\qb^2}{e^{\qb/T}+1}
\nonumber \\
&&~~~~~~~~~~
=48\alpha\zeta(3)\frac{\kb^2T^3}{\omega^4},
\label{eq65}
\end{eqnarray}
\noindent
where $\zeta(z)$ is the Riemann zeta function.

In the above, we have neglected by the imaginary part of ${\rm Im}\Delta_T\Pi_{00}$
which originates from the second line of Eq.~(\ref{eq41}).
It can be shown, however, that under the condition (\ref{eq57})
${\rm Im}\Delta_T\Pi_{00}$ is exponentially small.
Really, from the second line of Eq.~(\ref{eq41}) and the first formula of
Eq.~(\ref{eq43}) we have
\begin{equation}
{\rm Im}\Delta_T\Pi_{00}(\omega,\kb)=-\frac{8\alpha}{\v^2\tilde{p}(\omega,\kb)}
\int_{(\omega-\v\kb)/2}^{(\omega+\v\kb)/2}\!\!\!d\qb
\frac{\left[\v^2\kb^2-(2\qb-\omega)^2\right]^{1/2}}{e^{\qb/T}+1}.
\label{eq66}
\end{equation}
\noindent
Introducing the new variable $t=2\qb/(\v\kb)$, this can be represented
in the form
\begin{eqnarray}
&&
{\rm Im}\Delta_T\Pi_{00}(\omega,\kb)=-\frac{4\alpha\kb^2}{\tilde{p}(\omega,\kb)}
\int_{\omega/(\v\kb)-1}^{\omega/(\v\kb)+1}\!\!\!
\frac{dt}{e^{\v\kb t/(2T)}+1}
\nonumber \\
&&~~~~~~~~~~~~~~~\times
\sqrt{1-\left(t-\frac{\omega}{\v\kb}\right)^2}.
\label{eq67}
\end{eqnarray}
\noindent
Due to Eq.~(\ref{eq60}), this is an integration over the narrow interval centered
at $\omega/(\v\kb)$. At the ends of this interval the square root in the
integrand is equal to zero and in the middle to unity. Thus, the trapezium
estimation of the integral results in
\begin{equation}
{\rm Im}\Delta_T\Pi_{00}(\omega,\kb)\approx
-\frac{4\alpha\kb^2}{\sqrt{\omega^2-\v^2\kb^2}}
\,
\frac{1}{e^{\omega/(2T)}+1}
\label{eq68}
\end{equation}
\noindent
confirming an exponential smallness of the imaginary part of the polarization
tensor at high frequencies.

We are coming to calculation of the reflection coefficients and reflectivities
of graphene in the scattering region (\ref{eq28}). For photons on a mass-shell
the condition $\kb^2+k_3^2=\omega^2$ is satisfied, so that $\kb$ can be
expressed as
\begin{equation}
\kb=\omega\,\sin\theta_i,
\label{eq69}
\end{equation}
\noindent
where $\theta_i$ is the angle of incidence. From Eqs.~(\ref{eq37}) and (\ref{eq65})
we obtain the following expression for the 00-component of the polarization tensor
at high frequencies
\begin{equation}
\Pi_{00}(\omega,\kb)\approx
\frac{i\pi\alpha\kb^2}{\tilde{p}(\omega,\kb)}+48\alpha\zeta(3)
\frac{\kb^2T^3}{\omega^4}.
\label{eq70}
\end{equation}
\noindent
Substituting this in Eq.~(\ref{eq13}) for the TM reflection coefficient and using
Eq.~(\ref{eq69}) and the inequalities $\v\ll 1$, $T/\omega\ll 1$, one obtains
\begin{eqnarray}
&&
r_{\rm TM}(\omega,\theta_i)\approx\frac{\pi\alpha\cos\theta_i\left(1+\frac{\v^2}{2}\sin^2\theta_i\right)
-48i\alpha\zeta(3)\frac{T^3}{\omega^3}\cos\theta_i}{2+
\pi\alpha\cos\theta_i\left(1+\frac{\v^2}{2}\sin^2\theta_i\right)
-48i\alpha\zeta(3)\frac{T^3}{\omega^3}\cos\theta_i}
\nonumber \\
&&~~~~
\approx\frac{\pi\alpha\cos\theta_i}{2+\pi\alpha\cos\theta_i}\,
\left[1+\frac{\v^2}{2}\sin^2\theta_i-i\frac{48\zeta(3)}{\pi}
\left(\frac{T}{\omega}\right)^3\right].
\label{eq71}
\end{eqnarray}
\noindent
The reflectivity of the TM polarized light from graphene under the condition
(\ref{eq57}) follows from Eq.~(\ref{eq71})
\begin{equation}
|r_{\rm TM}(\omega,\theta_i)|^2
\approx\frac{\pi^2\alpha^2\cos^2\theta_i}{(2+\pi\alpha\cos\theta_i)^2}\,
\left[1+{\v^2}\sin^2\theta_i+\left(\frac{48\zeta(3)}{\pi}\right)^2
\left(\frac{T}{\omega}\right)^6\right].
\label{eq72}
\end{equation}
\noindent
Note that the corrections to unity in the square brackets of Eq.~(\ref{eq72})  are
very minor. This is seen from the fact that $\v^2=(\v/c)^2\approx 10^{-5}$ and
$T/\omega=k_BT/(\hbar\omega)\approx 10^{-2}$ at room temperature and optical
frequencies. Because of this, in the high frequency region the relative
thermal correction to the TM reflectivity of graphene at room temperature is
of order $10^{-12}$ (at zero temperature the result (\ref{eq72}) was obtained
in Ref.~\cite{20}).

In Fig.~3, we plot the TM reflectivity of graphene computed by Eq.~(\ref{eq72})
as a function of $\theta_i$ (the lower line). At the normal incidence
($\theta_i=0$) we have
\begin{equation}
|r_{\rm TM}(\omega,0)|^2\approx
\frac{\pi^2\alpha^2}{(2+\pi\alpha)^2}\approx 1.28\times 10^{-4}
\label{eq73}
\end{equation}
\noindent
in agreement with the result of Ref.~\cite{8} obtained numerically using the
local model for the conductivity of graphene. As is seen in Fig.~3, at high
frequencies the TM reflectivity of graphene is rather small and decreases
with increasing angle of incidence.

To consider the TE reflectivity of graphene at high frequencies, we need the
asymptotic expression for the quantity $\Delta_T\Pi$ defined in Eq.~(\ref{eq58}).
Substituting Eqs.~(\ref{eq61}) and (\ref{eq62}) in Eq.~(\ref{eq58}), one obtains
\begin{eqnarray}
&&
{\rm Re}\Delta_T\Pi(\omega,\kb)\approx-32\alpha\kb^2
\int_{0}^{\infty}\frac{d\qb}{e^{\qb/T}+1}\frac{\qb^2}{\omega^2-\qb^2}
\label{eq74} \\
&&~~~
\approx-32\alpha\frac{\kb^2}{\omega^2}
\int_{0}^{\infty}\frac{\qb^2d\qb}{e^{\qb/T}+1}
=-48\alpha\zeta(3)\kb^2\frac{T^3}{\omega^2}.
\nonumber
\end{eqnarray}
\noindent
Similar to the case of ${\rm Im}\Delta_T\Pi_{00}$, it is easily seen that
${\rm Im}\Delta_T\Pi$ is exponentially small.

As a result, the asymptotic expression for $\Pi$ at high frequencies is obtained
from Eqs.~(\ref{eq59}) and (\ref{eq74})
\begin{equation}
\Pi(\omega,\kb)\approx
-i\pi\alpha\kb^2\tilde{p}(\omega,\kb)-48\alpha\zeta(3)\kb^2
\frac{T^3}{\omega^2}.
\label{eq75}
\end{equation}
\noindent
We substitute this equation in the second line of  Eq.~(\ref{eq13}) and use
Eq.~(\ref{eq69}) and the smallness of our parameters $\v$ and $T/\omega$:
\begin{eqnarray}
&&
r_{\rm TE}(\omega,\theta_i)\approx-\frac{\pi\alpha\omega\left(1-\frac{\v^2}{2}\sin^2\theta_i\right)
-48i\alpha\zeta(3)\frac{T^3}{\omega^2}}{2\omega\cos\theta_i+
\pi\alpha\omega\left(1-\frac{\v^2}{2}\sin^2\theta_i\right)
-48i\alpha\zeta(3)\frac{T^3}{\omega^2}}
\nonumber \\
&&~~~~
\approx-\frac{\pi\alpha\left(1-\frac{\v^2}{2}\sin^2\theta_i\right)}{2\cos\theta_i+
\pi\alpha}\,
\left[1-i\frac{48\zeta(3)}{\pi}
\left(\frac{T}{\omega}\right)^3\right].
\label{eq76}
\end{eqnarray}
\noindent
The reflectivity of the TE polarized light from graphene at high frequencies
is given by
\begin{equation}
|r_{\rm TE}(\omega,\theta_i)|^2
\approx\frac{\pi^2\alpha^2}{(2\cos\theta_i+\pi\alpha)^2}\,
\left[1-{\v^2}\sin^2\theta_i+\left(\frac{48\zeta(3)}{\pi}\right)^2
\left(\frac{T}{\omega}\right)^6\right].
\label{eq77}
\end{equation}
\noindent
Similar to Eq.~(\ref{eq72}), the corrections to unity in the square brackets
of Eq.~(\ref{eq77}) are very small. At $T=0$ the result (\ref{eq77}) was
derived in Ref.~\cite{20}.

The TE reflectivity of graphene computed by Eq.~(\ref{eq77}) as a function of
$\theta_i$ is shown in Fig.~3 (the upper line).
At the normal incidence ($\theta_i=0$) we have
\begin{equation}
|r_{\rm TE}(\omega,0)|^2=|r_{\rm TM}(\omega,0)|^2,
\label{eq78}
\end{equation}
\noindent
where $|r_{\rm TM}(\omega,0)|^2$ is given in Eq.~(\ref{eq73}).
This is in agreement with Ref.~\cite{8}. However, with increasing
angle of incidence, $|r_{\rm TE}(\omega,\theta_i)|^2$ monotonously
increases, as opposite to  $|r_{\rm TM}(\omega,\theta_i)|^2$, which
decreases with the increase of $\theta_i$. Thus, at this point our
quantitative results disagree with the qualitative conclusion of
Ref.~\cite{8} that in the framework of the Dirac model
$|r_{\rm TE}(\theta_i)|^2=|r_{\rm TM}(0)|^2$   holds for any angle
of incidence.

\section{Reflectivity of graphene at low and intermediate frequencies}

In this section, we derive analytic asymptotic expressions for the polarization
tensor and reflectivities of graphene in the scattering frequency interval
(\ref{eq28}) at low frequencies satisfying the condition $\omega\ll T$.
We also perform numerical computations using the exact expressions for the
polarization tensor and find explicitly the application regions for the
asymptotic expressions at low and high frequencies.

We start from the imaginary part of the thermal correction $\Delta_T\Pi_{00}$
in the first formula of Eq.~(\ref{eq43}), which is determined by the second
line of Eq.~(\ref{eq41})
\begin{eqnarray}
&&
{\rm Im}\Delta_T\Pi_{00}(\omega,\kb)=-\frac{8\alpha}{\v^2\tilde{p}(\omega,\kb)}
\int_{(\omega-\v\kb)/2}^{(\omega+\v\kb)/2}\frac{d\qb}{e^{\qb/T}+1}
\sqrt{\v^2\kb^2-(\omega-2\qb)^2}
\label{eq79} \\
&&~~
=-\frac{8\alpha}{\v^2\tilde{p}(\omega,\kb)}\left[
\int_{(\omega-\v\kb)/2}^{\omega/2}\frac{d\qb}{e^{\qb/T}+1}
\sqrt{\v^2\kb^2-(\omega-2\qb)^2} \right.
\nonumber \\
&&~~~~~~~~~~~~~~~~~
\left.+
\int_{\omega/2}^{(\omega+\v\kb)/2}\!\!\frac{d\qb}{e^{\qb/T}+1}
\sqrt{\v^2\kb^2-(\omega-2\qb)^2} \right].
\nonumber
\end{eqnarray}
\noindent
It is convenient to introduce new variables $x=\omega-2\qb$ and $x=2\qb-\omega$
in the first and second integrals on the right-hand side of Eq.~({\ref{eq79}),
respectively. This results in
\begin{equation}
{\rm Im}\Delta_T\Pi_{00}(\omega,\kb)=-\frac{4\alpha}{\v^2\tilde{p}(\omega,\kb)}
\int_{0}^{\v\kb}dx\sqrt{\v^2\kb^2-x^2}\frac{2+
e^{\frac{\omega+x}{2T}}+e^{\frac{\omega-x}{2T}}}{\left(
e^{\frac{\omega-x}{2T}}+1\right)\left(e^{\frac{\omega+x}{2T}}+1\right)}.
\label{eq80}
\end{equation}
\noindent
Now we take into account that $x\ll\omega$ because in the scattering region
under consideration $\v\kb\ll\kb\leq\omega$. In the lowest order one can put
$\omega\pm x\approx\omega$ and using the condition $\omega\ll T$ obtain
\begin{equation}
{\rm Im}\Delta_T\Pi_{00}(\omega,\kb)\approx
-\pi\alpha\frac{\kb^2}{\tilde{p}(\omega,\kb)}\,\frac{2}{e^{\frac{\omega}{2T}}+1}.
\label{eq81}
\end{equation}
\noindent
Comparing this result with Eq.~(\ref{eq37}), we conclude that at low frequencies
the thermal correction to $\Pi_{00}$ gives the same contribution to the imaginary
part of $\Pi_{00}$ as the zero-temperature part $\Pi_{00}^{(0)}$.

The real part of the thermal correction $\Delta_T\Pi_{00}$ is determined by
Eq.(\ref{eq40}) and by the first and third lines of Eq.~(\ref{eq41}).
Substituting these in the first formula of Eq.~(\ref{eq43}), one finds
\begin{eqnarray}
&&
{\rm Re}\Delta_T\Pi_{00}(\omega,\kb)=\frac{16\alpha}{\v^2}
\left[\int_{0}^{\infty}\frac{d\qb}{e^{\qb/T}+1}-\frac{1}{2\tilde{p}(\omega,\kb)}
\int_{0}^{\infty}d\qb\frac{\sqrt{(\omega+2\qb)^2-\v^2\kb^2} }{e^{\qb/T}+1}
\right.
\nonumber \\
&&~~~
-\frac{1}{2\tilde{p}(\omega,\kb)}
\int_{0}^{(\omega-\v\kb)/2}d\qb\frac{\sqrt{(\omega-2\qb)^2-\v^2\kb^2} }{e^{\qb/T}+1}
\nonumber \\
&&~~~~~~~~~~~~
\left.
+\frac{1}{2\tilde{p}(\omega,\kb)}
\int_{(\omega+\v\kb)/2}^{\infty}d\qb\frac{\sqrt{(\omega-2\qb)^2-\v^2\kb^2} }{e^{\qb/T}+1}
\right].
\label{eq82}
\end{eqnarray}
\noindent
In this equation, we make the following identical transformations. The first integral is
calculated explicitly. In the second and third integrals we introduce the new variables
$y=\omega+2\qb$ and  $y=\omega-2\qb$, respectively. In the fourth integral we introduce
the new variable $y=2\qb-\omega$. Then Eq.~(\ref{eq82}) takes the form
\begin{eqnarray}
&&
{\rm Re}\Delta_T\Pi_{00}(\omega,\kb)=\frac{16\alpha}{\v^2}
\left[T\ln 2-\frac{1}{4\tilde{p}(\omega,\kb)}
\int_{\omega}^{\infty}\frac{dy}{e^{(y-\omega)/(2T)}+1}\sqrt{y^2-\v^2\kb^2}
\right.
\nonumber \\
&&~~~
-\frac{1}{4\tilde{p}(\omega,\kb)}
\int_{\v\kb}^{\omega}\frac{dy}{e^{(\omega-y)/(2T)}+1}\sqrt{y^2-\v^2\kb^2}
\nonumber \\
&&~~~~~~~~~~~~
\left.
+\frac{1}{4\tilde{p}(\omega,\kb)}
\int_{\v\kb}^{\infty}\frac{dy}{e^{(\omega+y)/(2T)}+1}\sqrt{y^2-\v^2\kb^2}
\right].
\label{eq83}
\end{eqnarray}
\noindent
In the last integral on the right-hand side of Eq.~(\ref{eq83}) let us separate the
integration region $[\v\kb,\infty)$ into two parts: $[\v\kb,\omega]$  and
$[\omega,\infty)$. Then we combine the integral over the first part with the
second integral on the right-hand side of Eq.~(\ref{eq83})  and the integral over
the second part --- with the first. As a result, we have
\begin{eqnarray}
&&
{\rm Re}\Delta_T\Pi_{00}(\omega,\kb)=\frac{16\alpha}{\v^2}
\left[T\ln 2 +
\frac{1}{4\tilde{p}(\omega,\kb)}
\int_{\v\kb}^{\omega}dy\sqrt{y^2-\v^2\kb^2}
\right.
\nonumber\\
&&
~~~~~~\times\left(
\frac{e^{-\omega/(2T)}}{e^{y/(2T)}+e^{-\omega/(2T)}}-
\frac{e^{y/(2T)}}{e^{y/(2T)}+e^{\omega/(2T)}}\right)
\label{eq84} \\
&&~~
\left.
-\frac{1}{4\tilde{p}(\omega,\kb)}
\int_{\omega}^{\infty}
dy\sqrt{y^2-\v^2\kb^2}
\left(
\frac{e^{\omega/(2T)}}{e^{y/(2T)}+e^{\omega/(2T)}}-
\frac{e^{-\omega/(2T)}}{e^{y/(2T)}+e^{-\omega/(2T)}}
\right)\right].
\nonumber
\end{eqnarray}
\noindent
This is an exact result which can be used both for numerical computations
and for obtaining the asymptotic expressions at low frequency.

It is easily seen that at $\omega\ll T$ the first integral
on the right-hand side of Eq.~(\ref{eq84}) is negligibly small.
To see this we expand the quantity in round brackets in porews
of small parameters $\omega/T$ and $y/T$ and find the leading order term
equal to $-y/(4T)$. After integration with respect to $y$ the magnitude
of the first integral in Eq.~(\ref{eq84}) together with its coefficient
appears to be $\v^2\kb^2/(48T)$.

The second integral on the right-hand side of Eq.~(\ref{eq84}) can be
calculated taking into account that $\omega\ll T$ and $\v\kb\ll\omega<y$.
Then we can expand in the small parameters $\v\kb/y$ and $\omega/T$.
Keeping in mind also that the main contribution to the integral is
given by $y\sim T$, we can replace the lower integration limit with zero.
Then from Eq.~(\ref{eq84}) we obtain
\begin{eqnarray}
&&
{\rm Re}\Delta_T\Pi_{00}(\omega,\kb)\approx\frac{16\alpha}{\v^2}
\left\{\vphantom{\frac{e^{y/(2T)}}{(e^{y/(2T)}+1)^2}}
T\ln 2 -
\frac{1}{4\omega}\left(1+\frac{\v^2\kb^2}{2\omega^2}\right)
\right.
\nonumber \\
&&~~~~
\left.\times
\frac{\omega}{T}
\int_{0}^{\infty}ydy\frac{e^{y/(2T)}}{[e^{y/(2T)}+1]^2}
\right\}
\label{eq85}
\end{eqnarray}
\noindent
and after integration by parts arrive at
\begin{equation}
{\rm Re}\Delta_T\Pi_{00}(\omega,\kb)\approx -8\alpha T\ln 2 \frac{\kb^2}{\omega^2}.
\label{eq86}
\end{equation}
\noindent
Thus, the contribution of the second integral in Eq.~(\ref{eq84}) is really much
larger than that of the first.

As a result, from Eqs.~(\ref{eq37}), (\ref{eq81}) and (\ref{eq86}) one obtains
the following expression for the 00-component of the polarization tensor at
low frequency
\begin{equation}
\Pi_{00}(\omega,\kb)=\frac{i\pi\alpha\omega\kb^2}{4T\tilde{p}(\omega,\kb)}-
 8\alpha T\ln 2 \frac{\kb^2}{\omega^2}.
\label{eq87}
\end{equation}
\noindent
Now we substitute Eq.~(\ref{eq87}) in Eq.~(\ref{eq13}), use  Eq.~(\ref{eq69})
and replace $\tilde{p}(\omega,\kb)$ with $\omega$. In the case of low frequencies
 this replacement is appropriate because the terms of order $\v^2$ are much
 smaller than the thermal correction. Then we have
\begin{equation}
r_{\rm TM}(\omega,\theta_i)\approx
\frac{\pi\alpha\frac{\omega}{8T}\cos\theta_i
+4i\alpha \ln 2\,\frac{T}{\omega}\,\cos\theta_i}{1+
\pi\alpha\frac{\omega}{8T}\cos\theta_i
+4i\alpha \ln2\,\frac{T}{\omega}\,\cos\theta_i}.
\label{eq88}
\end{equation}
\noindent
{}From Eq.~(\ref{eq88}) the reflectivity of the TM polarized light from
graphene at low frequencies is given by
\begin{equation}
|r_{\rm TM}(\omega,\theta_i)|^2\approx
\frac{\pi^2\alpha^2\frac{\omega^2}{64T^2}\cos^2\theta_i+16\alpha^2 \ln^2\!2\,\frac{T^2}{\omega^2}\,\cos^2\theta_i}{\left(1+
\pi\alpha\frac{\omega}{8T}\cos\theta_i\right)^2+
16\alpha^2 \ln^2\!2\,\frac{T^2}{\omega^2}\,\cos^2\theta_i}.
\label{eq89}
\end{equation}
\noindent
Note that at low frequency both real and imaginary parts of the $\Pi_{00}$ component
(\ref{eq87}) contribute to the reflection coefficient (\ref{eq88}) and reflectivity
(\ref{eq89}). The thermal correction contributes essentially to both quantities.
At the normal incidence we have
\begin{equation}
|r_{\rm TM}(\omega,0)|^2\approx
\frac{\pi^2\alpha^2\frac{\omega^2}{64T^2}+16\alpha^2 \ln^2\!2\,\frac{T^2}{\omega^2}}{\left(1+
\pi\alpha\frac{\omega}{8T}\right)^2+16\alpha^2 \ln^2\!2\,\frac{T^2}{\omega^2}}.
\label{eq90}
\end{equation}

In Fig.~4, we plot the reflectivity of graphene (\ref{eq90}) at $\theta_i=0$ as
a function of frequency measured in K at three different temperatures $T=10\,$K,
100\,K, and 300\,K (the lower, middle, and upper lines, respectively).
As is seen in Fig.~4, at $\omega=0$ the reflectivity is equal to unity and decreases
monotonously with increasing frequency. The comparison with the results of
numerical computations (see below) shows that even at the lowest considered
temperature $T=10\,$K the asymptotic expression (\ref{eq90}) is rather precise.
Our analytic results of Fig~4 obtained using the polarization tensor are in rather
good agreement with the computational results of Ref.~\cite{7} (see their Fig.~2, left)
found using the local model for the conductivity of graphene. Note, however, that
with decreasing temperature the local model supposedly overestimates the TM
reflectivity of graphene.

We are coming now to the calculation of the TE reflectivity of graphene at low
frequency. Besides $\Pi_{00}$, the reflection coefficient $r_{\rm TE}$ in
Eq.~(\ref{eq13}) is expressed via the quantity $\Pi$ defined in Eqs.~(\ref{eq58})
and (\ref{eq59}). To calculate $\Delta_T\Pi$ at $\omega\ll T$, it is convenient to
use its definition in the first line of Eq.~(\ref{eq58}) and already obtained
results (\ref{eq81}) and (\ref{eq86}) for $\Delta_T\Pi_{00}$. The quantity
$\Delta_T\Pi_{\rm tr}$ is given in the second formula of Eq.~(\ref{eq43}).
Both its imaginary and real parts at low frequency are calculated using the same
procedure, as was applied above in this section to calculate
${\rm Im}\Delta_T\Pi_{00}$ and  ${\rm Re}\Delta_T\Pi_{00}$. The results are
\begin{eqnarray}
&&
{\rm Im}\Delta_T\Pi_{\rm tr}(\omega,\kb)\approx \pi\alpha\,\frac{p^2(\omega,\kb)+
\tilde{p}^{\,2}(\omega,\kb)}{\tilde{p}(\omega,\kb)}\,
\frac{2}{e^{\frac{\omega}{2T}}+1},
\label{eq91} \\
&&
{\rm Re}\Delta_T\Pi_{\rm tr}(\omega,\kb)\approx 8\alpha T\ln\!2\,\frac{p^2(\omega,\kb)+
\tilde{p}^{\,2}(\omega,\kb)+2\v^2\kb^2}{\omega^2}.
\nonumber
\end{eqnarray}
\noindent
Then from the second line of Eq.~(\ref{eq37}) and Eq.~(\ref{eq91}) one finds
\begin{eqnarray}
&&
\Pi_{\rm tr}\approx 8\alpha T\ln\!2\,\frac{p^2(\omega,\kb)+
\tilde{p}^{\,2}(\omega,\kb)+2\v^2\kb^2}{\omega^2}
\nonumber \\
&&~~~~~~~~~~
+i\frac{\omega}{4T}
\pi\alpha\,\frac{p^2(\omega,\kb)+
\tilde{p}^{\,2}(\omega,\kb)}{\tilde{p}(\omega,\kb)}.
\label{eq92}
\end{eqnarray}
\noindent
{}Finally, from Eqs.~(\ref{eq87}) and (\ref{eq92}) we arrive at
\begin{equation}
\Pi(\omega,\kb) \approx 8\alpha T\ln\!2\,\kb^2-i\frac{\omega}{4T}
\pi\alpha\kb^2\tilde{p}(\omega,\kb),
\label{eq93}
\end{equation}
\noindent
where we have taken into account that $\v^2\kb^2\ll\omega^2$.

Substituting Eq.~(\ref{eq93}) in Eq.~(\ref{eq13}) for the TE reflection
coefficient, using Eq.~(\ref{eq69}) and taking into account that
$\tilde{p}(\omega,\kb)\approx\omega$ in the scattering frequency interval,
one obtains
\begin{equation}
r_{\rm TE}(\omega,\theta_i)=-\frac{\pi\alpha\frac{\omega}{8T}+4i\alpha\ln\!2\,
\frac{T}{\omega}}{\cos\theta_i+\pi\alpha\frac{\omega}{8T}+4i\alpha\ln\!2\,
\frac{T}{\omega}}\, .
\label{eq94}
\end{equation}
\noindent
The reflectivity of the TE polarized light from graphene at low frequencies
is given by
\begin{equation}
|r_{\rm TE}(\omega,\theta_i)|^2=\frac{\pi^2\alpha^2\frac{\omega^2}{64T^2}+16\alpha^2\ln^2\!2\,
\frac{T^2}{\omega^2}}{\left(\cos\theta_i+\pi\alpha
\frac{\omega}{8T}\right)^2+16\alpha^2\ln^2\!2\,
\frac{T^2}{\omega^2}}\, .
\label{eq95}
\end{equation}
\noindent
{}From the comparison of Eq.~(\ref{eq95}) with Eq.~(\ref{eq90}) it is seen that
at the normal incidence
\begin{equation}
|r_{\rm TE}(\omega,0)|^2=|r_{\rm TM}(\omega,0)|^2,
\label{eq96}
\end{equation}
\noindent
as it should be, where the right-hand side of this equation is given by Eq.~(\ref{eq90}).
Thus, Fig.~4 demonstrating dependence of the reflectivity of graphene on frequency at the
normal incidence is equally applicable to both TM and TE polarized light.

Now we illustrate dependence of the reflectivities of graphene at low frequency on the
angle of incidence. Calculations were performed using Eqs.~(\ref{eq89}) and (\ref{eq95})
for the TM and TE polarized light, respectively. In Fig.~5 the calculation results are
presented by the upper pair of solid lines computed at $\omega/T=0.01$ and lower pair
corresponding to $\omega/T=0.03$. As is seen from Fig.~5, at low frequencies the
reflectivities of TM polarized light decrease with increasing $\theta_i$, whereas
the reflectivities of TE polarized light are the monotonously increasing functions
of $\theta_i$. In this sense the dependence of reflectivities of graphene at low
frequencies on the angle of incidence is similar to that at high frequencies.
The single difference is that the magnitudes of reflectivity at low frequencies
are several orders of magnitude higher than at high frequencies.

In the end of this section we present the results of numerical computations for the
reflectivity of graphene over a wide range of frequencies and compare them with
the analytic asymptotic results obtained in this and previous sections.
As an example, we choose the incidence angle $\theta_i=30^{\circ}$ and perform
numerical computations of the TM and TE reflectivities as functions of
$\omega/T$ in the wide range from $10^{-3}$ to $10^3$. Let us start from the
TM polarized light. In this case computations are done using Eqs.~(\ref{eq13}),
(\ref{eq18}), (\ref{eq41}) and (\ref{eq43}). The computational results are
shown by the solid line in Fig.~6(a) plotted in the double logarithmic scale.
In the same figure, the long-dashed line shows the analytic asymptotic results
at high frequencies. They are calculated using Eq.~(\ref{eq72}) at
$\theta=30^{\circ}$. The analytic asymptotic results at low frequencies are
calculated by Eq.~(\ref{eq89}) and shown by the short-dashed line.
As is seen in Fig.~6(a), the analytic asymptotic results are in a very good
agreement with the results of numerical computations over a wide frequency
region. Thus, at low frequencies the relative deviation of the short-dashed
line less than 5\% occurs at all $\omega/T<0.3$ [$\log(\omega/T)<-0.52$].
At high frequencies less than 5\% relative deviation holds at all
$\omega/T>9$ [$\log(\omega/T)>0.95$].

Finally we consider the TE polarized light incident on graphene at the same
angle $\theta_i=30^{\circ}$. We perform numerical computations using Eqs.~(\ref{eq13}),
(\ref{eq41}), (\ref{eq58}) and (\ref{eq59}).
In Fig.~6(b), the computational results are
presented by the solid line plotted in the double logarithmic scale.
Similar to Fig.~6(a), the analytic results of Eq.~(\ref{eq77}) at high frequencies
are shown by the long-dashed line and
the analytic results of Eq.~(\ref{eq95}) at low frequencies
are shown by the short-dashed line. From Fig.~6(b) it is seen that at low
frequencies less than 5\% relative deviation
occurs at $\omega/T<0.25$ [$\log(\omega/T)<-0.6$], and at
high frequencies less than 5\% relative deviation holds at all
$\omega/T>9$ [$\log(\omega/T)>0.95$].
This solves the question when one can use simple asymptotic formulas and when more
cumbersome numerical computations are necessary.

\section{Conclusions and discussion}

In the foregoing, we have developed quantum field theoretical description for the
reflectivity of graphene described in the framework of the Dirac model.
This description exploits the polarization tensor of graphene at nonzero temperature
in (2+1)-dimensional space-time. The previously known representation of this tensor
\cite{10} obtained in thermal quantum field theory in Matsubara formulation is
applicable only at the imaginary Matsubara frequencies and does not admit physically
reasonable analytic continuation to the real frequency axis. Here we derive another
representation for the polarization tensor of graphene which possesses the
required analytic properties over the entire plane of complex frequencies.
At zero temperature our polarization tensor coincides with that obtained in
Ref.~\cite{9}. At nonzero temperature our tensor takes the same values as the one
obtained in Ref.~\cite{10} at all Matsubara frequencies. This justifies all
numerous applications of the polarization tensor of Ref.~\cite{10} to calculation
of the Casimir interaction in systems including graphene and graphene-coated
substrates \cite{10,11,12,13,14,15,16,17}.
However, as shown in this paper, the polarization tensor of Ref.~\cite{10} is
an oscillating function along the imaginary frequency axis, and this precludes
the use of its analytic continuation to real frequency axis for theoretical
description of the reflectivity of graphene. On the contrary, in the
representation obtained here the polarization tensor is a monotonously
decreasing function of the imaginary frequency and admits the analytic continuation
to real frequency axis satisfying all physical requirements.

Using the derived representation for the polarization tensor at nonzero temperature,
we have obtained analytic expressions for the reflection coefficients and
reflectivities of graphene for two independent polarizations of the electromagnetic
field in the asymptotic regions of high and low frequencies, as compared to
temperature. For the transverse magnetic ($p$ polarized) light our results are in
a good agreement with those previously obtained at high \cite{8} and low \cite{7}
frequencies using the local model for the conductivity of graphene. However, some
qualitative conclusions about the reflectivity of TE ($s$ polarized) light made
previously are not supported by our exact calculations.
We have investigated dependences of the TM and TE reflectivities of graphene
as functions of frequency and incidence angle. In the latter case it was shown
that the TM reflectivity is a decreasing function of the incidence angle, whereas
the TE reflectivity increases with increase of the angle of incidence.

Furthermore, we have performed numerical computations of the TM and TE
reflectivities of graphene over the wide frequency range. By comparing the
analytic and computational results the application regions for the asymptotic
expansions at low and high frequencies were determined. It was shown that
simple analytic expressions for the reflectivity of graphene obtained here produce
very accurate results in the wide regions of frequencies and temperatures.

By and large the developed formalism solves the problem of the reflectivity of
pristine (gapless) graphene and graphene characterized by some nonzero mass-gap
parameter. In future it would be interesting to generalize this formalism to the
case of nonzero chemical potential and consider applications to doped graphene
sheets.

\appendix
\section{Polarization tensor in (2+1)-dimensions at nonzero temperature}
In this Appendix, we display the details for the calculation of the polarization
tensor of graphene at nonzero temperature.
By making the replacement (\ref{eq6}) in Eq.~(\ref{eq4}), we represent the polarization
tensor in the form
\be
\Pi^{\mu\nu}(i\xi_l,{\k})=
\Pi^{\mu\nu}(k)=-8\pi\alpha T
\sum\limits_{n=-\infty}^{\infty}
\int\frac{d{\q}_{\bot}}{(2\pi)^2}\, {\rm tr}\, S(q)\tilde{\gamma}^\mu S(q-k)\tilde{\gamma}^\nu,
\label{A.1}\ee
where the spinor propagator is given by
\be S(q)=\frac{1}{i\slashed q-m}=\frac{i\slashed q+m}{q^2-m^2-i0}
\label{A.2}\ee
and the notations introduced in Eqs.~(\ref{eq5}) and (\ref{eq6}) are used here and
below.

By calculating the trace over the $\gamma$-matrices, we get
\be
\Pi^{\mu\nu}(k)=-\frac{32\pi\alpha}{\v^2} T
\sum\limits_{n=-\infty}^{\infty}
\int\frac{d{\q}_{\bot}}{(2\pi)^2}\, \frac{Z^{\mu\nu}(k,q)}{R(k,q)},
\label{A.3}\ee
where
\begin{eqnarray}
&&
Z^{\mu\nu}(k,q)=\eta_{\mu^{\prime}}^{\mu}\eta_{\nu^{\prime}}^{\nu}
\left[q^{\mu^{\prime}}(q^{\mu^{\prime}}-\tilde{k}^{\,\nu^{\prime}})+
(q^{\mu^{\prime}}-\tilde{k}^{\,\mu^{\prime}})q^{\nu^{\prime}}
\right.
\nonumber \\
&&~~~~~~~~~~~~\left.
-
q_{\alpha}(q^{\alpha}-\tilde{k}^{\,\alpha})
g^{\mu^{\prime}\nu^{\prime}}+
m^2g^{\mu^{\prime}\nu^{\prime}}\right],
\nonumber \\
&&
R(k,q)=(q_{\mu}q^{\mu}-m^2+i0)\left[
(q_{\mu}-\tilde{k}_{\mu})(q^{\mu}-\tilde{k}^{\,\mu})-m^2+i0\right].
\label{A.4}
\end{eqnarray}
\noindent
In there equations, $\tilde{k}^{\,\mu}\equiv (ik_{l4},\v{\k})$ and $g^{\mu\nu}$
is the metric tensor in (2+1)-dimensions with
$g^{00}=-g^{11}=-g^{22}=1$. Note also that we have made the substitution
${\q}_{\bot}\to{\q}_{\bot}/\v$ in the integration in Eq.~(\ref{A.1}).

For the independent quantities (\ref{eq8}) characterizing the polarization tensor
at any temperature it follows
\be
\Pi_{00({\rm tr})}(k)=-\frac{32\pi\alpha}{\v^2} T
\sum\limits_{n=-\infty}^{\infty}
\int\frac{d{\q}_{\bot}}{(2\pi)^2}\, \frac{Z_{00({\rm tr})}(k,q)}{R(k,q)}.
\label{A.5}\ee
Here, by using Eq.~(\ref{A.4}), we obtain
\begin{eqnarray}
&&
Z_{00}(k,q)=-q_{4n}(q_{4n}-k_{4l})+{\q}_{\bot}
({\q}_{\bot}-\v\k)+m^2,
\nonumber \\
&&
Z_{\rm tr}(k,q)=-(1-2\v^2)q_{4n}(q_{4n}-k_{4l})+{\q}_{\bot}
({\q}_{\bot}-\v\k)
\nonumber \\
&&~~~~~~~~~~~~
+(1+2\v^2)m^2,
\label{A.6} \\
&&
R(q,k)=[q_{4n}^2+\Gamma^2(\qb)]\left[(q_{4n}^2-k_{4l}^2)+
\tilde{\Gamma}^2(\k,{\q}_{\bot})\right],
\nonumber
\end{eqnarray}
\noindent
where $\Gamma(\qb)$ is defined in Eq.~(\ref{eq21}) and
\begin{equation}
\tilde{\Gamma}(\k,{\q}_{\bot})=\left[({\q}_{\bot}-\v\k)^2+m^2
\right]^{1/2}.
\label{A.7}
\end{equation}
\noindent
We mention that $\Pi_{00}$ in (\ref{A.5}) coincides up to an overall factor
with Eq.~(A3) in Ref.~\cite{22}.

Now, following the idea behind the Abel-Plana formula
(see, e.g., Ref.~\cite{25}),
we represent the Matsubara sum over $n$ in
Eq.~(\ref{A.5}) as an integral in the complex $q_4$ plane,
\be
\Pi_{00({\rm tr})}(k)=\frac{32\pi\alpha}{\v^2}\!\!
\int\limits_{\gamma_1\cup\gamma_2}\frac{dq_4}{2\pi}
\int\frac{d{\q}_{\bot}}{(2\pi)^2}\,
\frac{1}{e^{iq_4/T}+1}\,
\frac{Z_{00({\rm tr})}(k,q)}{R(k,q)},
\label{A.8}\ee
where the integration path $\gamma_1\cup\gamma_2$ encircles the poles of
the real axis at the fermionic Matsubara frequencies indicated by dots
(see Fig.~\ref{last}).
In the same figure, the zeros of the quantity $R(q,k)$ in the denominator
of (\ref{A.8}) at $q_4=\pm i\Gamma$ and $q_4=\pm i\tilde{\Gamma}+k_{4l}$ are
shown as crosses.

It is convenient to represent Eq.~(\ref{A.8}) as the sum of two integrals.
The first of them is along the path $\gamma_1$ and the second is along the
path $\gamma_2$. In order to have a decreasing function in the upper half
plane under the integral along the path $\gamma_1$, we use the identity
\begin{equation}
\frac{1}{e^{iq_4/T}+1}=1-\frac{1}{e^{-iq_4/T}+1}.
\label{A.9}\ee
We substitute Eq.~(\ref{A.9}) in the first integral of Eq.~(\ref{A.8})
along the path $\gamma_1$. In so doing the unity results in the polarization
tensor at zero temperature
\begin{eqnarray}
&&
\Pi_{00({\rm tr})}^{(0)}(k)=\frac{32\pi\alpha}{\v^2}\!\!
\int\limits_{\gamma_1}\frac{dq_4}{2\pi}
\int\frac{d{\q}_{\bot}}{(2\pi)^2}\,
\frac{Z_{00({\rm tr})}(k,q)}{R(k,q)}
\label{A.10} \\
&&~~~~
=-\frac{32\pi\alpha}{\v^2}\!\!
\int_{-\infty}^{\infty}\frac{dq_4}{2\pi}
\int\frac{d{\q}_{\bot}}{(2\pi)^2}\,
\frac{Z_{00({\rm tr})}(k,q)}{R(k,q)}.
\nonumber
\end{eqnarray}
\noindent
The substitution of the second term of Eq.~(\ref{A.9}) in the first integral of Eq.~(\ref{A.8})
along the path $\gamma_1$ together with the second integral of Eq.~(\ref{A.8})
along the path $\gamma_2$ result in the thermal correction to the polarization tensor
\begin{eqnarray}
&&
\Delta_T\Pi_{00({\rm tr})}(k)=-\frac{32\pi\alpha}{\v^2}\!\!
\int\limits_{\gamma_1}\frac{dq_4}{2\pi}
\int\frac{d{\q}_{\bot}}{(2\pi)^2}\,
\frac{1}{e^{-iq_4/T}+1}\,
\frac{Z_{00({\rm tr})}(k,q)}{R(k,q)}
\nonumber \\
&&~~~~~~
+\frac{32\pi\alpha}{\v^2}\!\!
\int\limits_{\gamma_2}\frac{dq_4}{2\pi}
\int\frac{d{\q}_{\bot}}{(2\pi)^2}\,
\frac{1}{e^{iq_4/T}+1}\,
\frac{Z_{00({\rm tr})}(k,q)}{R(k,q)}.
\label{A.11}
\end{eqnarray}

It is well known that the polarization tensor has no ultraviolet divergencies
in (2+1)-dimensions. In fact, its divergence degree is +1, which would correspond
to a linear divergence. The latter, however, is not present due to the gauge
invariance. Otherwise, the linear divergence would reside in the zero-temperature
part $\Pi_{00({\rm tr})}^{(0)}$. The thermal corrections $\Delta_T\Pi_{00({\rm tr})}$
do not have any ultraviolet divergencies, which is evident due to the Boltzmann
factors.

Calculation of the polarization tensor at zero temperature (\ref{A.10}) was
performed in Ref.~(\ref{eq9}) with the result (\ref{eq18}).
Now we calculate the thermal correction to it (\ref{A.11}).
For this purpose, we close the integration pathes $\gamma_1$ and $\gamma_2$ by
the half circles of infinitely large radii in the upper and lower half planes,
respectively. These circles do not contribute to the result which is given
by the corresponding residua in the poles of both integrands shown in
Fig.~\ref{last} as crosses
\begin{eqnarray}
&&
\Delta_T\Pi_{00({\rm tr})}(k)=\frac{32\pi\alpha}{\v^2}\!\!
\int\frac{d{\q}_{\bot}}{(2\pi)^2}\,
\sum\limits_{\lambda=\pm 1}\left\{
\vphantom{\frac{Z_{00({\rm tr})}(k,k_{4l}+i\lambda\tilde{\Gamma}(\qb),\q_{\bot})}{2\lambda
\tilde{\Gamma}(\k,\q_{\bot})
[(i\lambda\tilde{\Gamma}(\k,\q_{\bot})+k_{4l})^2+{\Gamma}^{\,2}(\qb)]}}
\frac{1}{e^{\Gamma(\qb)/T}+1}\right.
\nonumber \\
&&~~~
\frac{Z_{00({\rm tr})}(k,i\lambda\Gamma(\qb),\q_{\bot})}{2\lambda\Gamma(\qb)
[(i\lambda\Gamma(\qb)-k_{4l})^2+\tilde{\Gamma}^{\,2}(\k,\q_{\bot})]}
\label{A.12}\\
&&~~~
\left.
+\frac{1}{e^{[-i\lambda k_{4l}+\tilde{\Gamma}({\scriptsize\k},\q_{\bot})/T}+1}\,
\frac{Z_{00({\rm tr})}(k,k_{4l}+i\lambda\tilde{\Gamma}(\k,\q_{\bot}),\q_{\bot})}{2\lambda
\tilde{\Gamma}(\k,\q_{\bot})
[(i\lambda\tilde{\Gamma}(\k,\q_{\bot})+k_{4l})^2+{\Gamma}^{\,2}(\qb)]}
\right\},
\nonumber
\end{eqnarray}
\noindent
where the contributions with $\lambda=1$ come from the upper half plane and those
with $\lambda=-1$ from the lower half plane.

Equation (\ref{A.12}) can be simplified taking into account that according to
Eq.~(\ref{eq6a}) $k_{4l}$ are the Matsubara frequencies and, thus,
$\exp(-i\lambda k_{4l}/T)=1$.
In fact doing just this way is necessary for a correct analytic continuation
to non-Matsubara frequencies since otherwise the replacement
$k_{4l}\to i\omega$ would not result in a function (\ref{A.12}) decreasing for
$\omega\to\pm\infty$ as is required for the analytic continuation.
In application to temperature Green functions this was first mentioned in
Ref.~\cite{24} and extensively elaborated in Ref.~\cite{23}.
A recent, mathematically equivalent, treatment can be found in
Ref.~\cite{22} [see Eq.~(A5) there].
Further we use the symmetry of the integrand under the substitution
$\q_{\bot}\to \v\k-\q_{\bot}$ and making the replacement $\lambda\to -\lambda$
in the second term of Eq.~(\ref{A.12}) we arrive at
\begin{eqnarray}
&&
\Delta_T\Pi_{00({\rm tr})}(k)=\frac{16\pi\alpha}{\v^2}\!\!
\int\frac{d{\q}_{\bot}}{(2\pi)^2}\,
\frac{1}{\Gamma(\qb)}\,
\frac{1}{e^{\Gamma(\qb)/T}+1}
\label{A.13}\\
&&~~\times
\sum\limits_{\lambda=\pm 1}
\frac{Z_{00({\rm tr})}(k,i\lambda\Gamma(\qb),\q_{\bot})+
Z_{00({\rm tr})}(k,k_{4l}-i\lambda\Gamma(\qb),\v\k-\q_{\bot})}{[
k_{4l}-i\lambda\Gamma(\qb)]^2+\tilde{\Gamma}^{\,2}(\k,\q_{\bot})}.
\nonumber
\end{eqnarray}
\noindent
As said above,
in this representation, the analytic continuation of $\Delta_T\Pi_{00({\rm tr})}$
from the Matsubara frequencies $k_{4l}$ to any frequency $k_4\equiv\omega$,
including the real frequency axis, can be done directly.

Using the polar coordinates ($\qb=|\q_{\bot}|=\sqrt{q_1^2+q_2^2},\varphi$) in the
$\q_{\bot}$-plane, it is possible to perform one of the two integrations in Eq.~(\ref{A.13}).
Taking into account that
\begin{equation}
2\q_{\bot}\k=2\qb\kb\cos{\varphi},
\label{A.14}
\end{equation}
\noindent
for the quantity in the denominator of (\ref{A.13}) we obtain
\begin{equation}
[k_{4l}-i\lambda\Gamma(\qb)]^2+\tilde{\Gamma}^{\,2}(\k,\q_{\bot})=
Q(k_{4l},\kb,\qb)-2\v\qb\kb\cos{\varphi}
\label{A.15}
\end{equation}
\noindent
with the notation
\begin{equation}
Q(k_{4l},\kb,\qb)=k_{4l}^2+\v^2\kb^2-2i\lambda k_{4l}\Gamma(\qb).
\label{A.16}
\end{equation}
\noindent
In the numerator of Eq.~{\ref{A.13}) we get
\begin{eqnarray}
&&
Z_{00({\rm tr})}(k,i\lambda\Gamma(\qb),\q_{\bot})+
Z_{00({\rm tr})}(k,k_{4l}-i\lambda\Gamma(\qb),\v\k-\q_{\bot})
\nonumber \\
&&~~~
=Q(k_{4l},\kb,\qb)-
2\v\qb\kb\cos{\varphi}+M_{00({\rm tr})}(k_{4l},\kb,\qb)
\label{A.17}
\end{eqnarray}
\noindent
with
\begin{eqnarray}
&&
M_{00}(k_{4l},\kb,\qb)=-(k_{4l}^2+\v^2\kb^2)+4i\lambda k_{4l}\Gamma(\qb)
+4\Gamma^{\,2}(\qb),
\nonumber \\
&&
M_{\rm tr}(k_{4l},\kb,\qb)=-(k_{4l}^2+\v^2\kb^2)+4\v^2m^2
\label{A.18} \\
&&~~~
+4(1-\v^2)i\lambda k_{4l}\Gamma(\qb)
+4(1-\v^2)\Gamma^{\,2}(\qb).
\nonumber
\end{eqnarray}

Rewriting the integral (\ref{A.13}) in the polar coordinates and using
Eqs.~(\ref{A.15}) and (\ref{A.17}), we obtain
\begin{eqnarray}
&&
\Delta_T\Pi_{00({\rm tr})}(k)=\frac{16\pi\alpha}{\v^2}\!\!
\int_{0}^{\infty}d\qb
\frac{\qb}{\Gamma(\qb)}\,
\frac{1}{e^{\Gamma(\qb)/T}+1}
\label{A.19}\\
&&~~\times\left[1+
\sum\limits_{\lambda=\pm 1}
\int_{0}^{2\pi}\frac{d\varphi}{2\pi}
\frac{M_{00({\rm tr})}(k_{4l},\kb,\qb)}{Q(k_{4l},\kb,\qb)-2\v\qb\kb\cos\varphi}
\right].
\nonumber
\end{eqnarray}
\noindent
The remaining angular integral can be carried out by using the formula
\begin{equation}
\int_{0}^{2\pi}\frac{d\varphi}{2\pi}\,
\frac{1}{Q-a\cos\varphi}=\frac{1}{\sqrt{Q^2-a^2}}
\label{A.20}
\end{equation}
\noindent
with $a\equiv 2\v\qb\kb$. Here, the branch of the square root must be
defined such that the sign of the imaginary part of the root
$\sqrt{Q^2-a^2}$ be the same as that of $Q$. This prescription holds for
$Q$ as given by Eq.~(\ref{A.16}) with real $k_{4l}$.
For a complex $k_4$ one needs to use the corresponding analytic continuation.
Substituting Eq.~(\ref{A.20}) in Eq.~(\ref{A.19}) and using the notation
\begin{equation}
N(k_{4l},\kb,\qb)=[Q^2(k_{4l},\kb,\qb)- (2\v\qb\kb)^2]^{1/2},
\label{A.21}
\end{equation}
\noindent
one arrives to the final result (\ref{eq20}) for the polarization tensor of graphene
defined at the imaginary Matsubara frequencies. The analytic continuation of this
tensor to the real frequency axis is described in Sec.~II.


\begingroup
\squeezetable
\begin{table}
\caption{The normalized values of the thermal correction to the
00-component of polarization tensor of graphene computed at the
Matsubara frequencies with different numbers (column 1) using the
representation of Ref.~\cite{10} (column 2) and in this paper
(column 3) at $\kb=10\xi_1$. The relative thermal correction is
given in column 4.}
\begin{ruledtabular}
\begin{tabular}{rlll}
$l$&$~~~~~\Delta_T\Pi_{00}^{(a)}/C$ & $~~~~~\Delta_T\Pi_{00}/C$ &
$\!\!\!\!\!\!\!\!\Delta_T\Pi_{00}/{\Pi_{00}^{(0)}}\,$(\%) \\
\hline
1 & $5.60604848904\times 10^{-5}$ & $5.60604848906\times 10^{-5}$ & 4.09 \\
2 & $5.0366515197\times 10^{-6}$ &$5.0366515198\times 10^{-6}$ & 0.73 \\
3 & $1.1114739021\times 10^{-6}$ & $1.1114739022\times 10^{-6}$ & 0.24 \\
4 & $3.6872960104\times 10^{-7}$ &$3.6872960109\times 10^{-7}$ & 0.108 \\
5 & $1.547382154\times 10^{-7}$ & $1.547382155\times 10^{-7}$ & 0.056 \\
6 & $7.56717034\times 10^{-8}$ & $7.56717035\times 10^{-8}$ & 0.033 \\
7 & $4.12036077\times 10^{-8}$ & $4.12036078\times 10^{-8}$ & 0.021 \\
8 & $2.42932778\times 10^{-8}$ & $2.42932779\times 10^{-8}$ & 0.014 \\
9 & $1.52276095\times 10^{-8}$ & $1.52276096\times 10^{-8}$ & 0.010 \\
10 & $1.00201160\times 10^{-8}$ &  $1.00201159\times 10^{-8}$ & 0.0073\\
11 & $6.85884188\times 10^{-9}$ & $6.85884190\times 10^{-9}$ & 0.0055
\end{tabular}
\end{ruledtabular}
\end{table}
\endgroup
\begingroup
\squeezetable
\begin{table}
\caption{The normalized values of the thermal correction (\ref{eq45}) to the
combination of the components of polarization tensor
of graphene computed at the
Matsubara frequencies with different numbers (column 1) using the
representation of Ref.~\cite{10} (column 2) and in this paper
(column 3) at $\kb=10\xi_1$. The relative thermal correction is
given in column 4.}
\begin{ruledtabular}
\begin{tabular}{rlll}
$l$&$~~~~~\Delta_T\Pi^{(a)}/D$ & $~~~~~\Delta_T\Pi/D$ &
$\!\!\!\!\!\!\!\!\Delta_T\Pi/{\Pi^{(0)}}\,$(\%) \\
\hline
1 & $1.12028612025\times 10^{-4}$ & $1.12028612024\times 10^{-4}$ & 4.08\\
2 & $4.02785688588\times 10^{-5}$ & $4.02785688597\times 10^{-5}$ & 0.73 \\
3 & $2.0002626999\times 10^{-5}$ & $2.0002626997\times 10^{-5}$ & 0.24 \\
4 & $1.1797934387\times 10^{-5}$ & $1.1797934384\times 10^{-5}$ & 0.108 \\
5 & $7.736289303\times 10^{-6}$ & $7.736289297\times 10^{-6}$ & 0.056 \\
6 & $5.44805008\times 10^{-6}$ & $5.44805006\times 10^{-6}$ & 0.033 \\
7 & $4.03778027\times 10^{-6}$ & $4.03778024\times 10^{-6}$ & 0.021 \\
8 & $3.10943615\times 10^{-6}$ & $3.10943614\times 10^{-6}$ & 0.014 \\
9 & $2.466807364\times 10^{-6}$ & $2.466807362\times 10^{-6}$ & 0.010 \\
10& $2.00397985\times 10^{-6}$ &  $2.00397988\times 10^{-6}$ & 0.0073 \\
11& $1.65980994\times 10^{-6}$ & $1.65980992\times 10^{-6}$ & 0.0055
\end{tabular}
\end{ruledtabular}
\end{table}
\endgroup
\begin{figure}[b]
\vspace*{-1cm}
\centerline{\hspace*{1cm}
\includegraphics{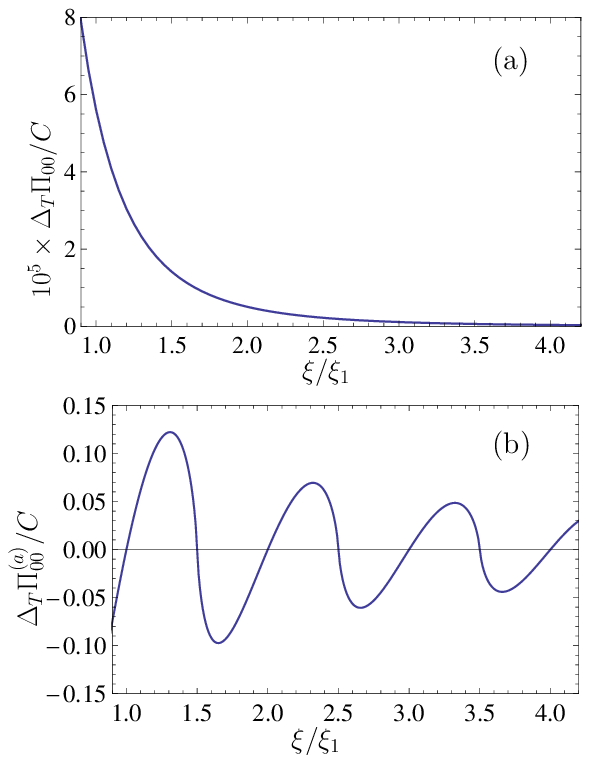}
}
\vspace*{-12cm}
\caption{\label{fg1}(Color online)
The normalized values of the thermal correction to the 00-component
of the polarization tensor of graphene computed as functions of imaginary
frequency at $\kb=10\xi_1$ (a) in this paper and (b) using the representation
of Ref.~\cite{10}.
}
\end{figure}

\begin{figure}[b]
\vspace*{-1cm}
\centerline{\hspace*{1cm}
\includegraphics{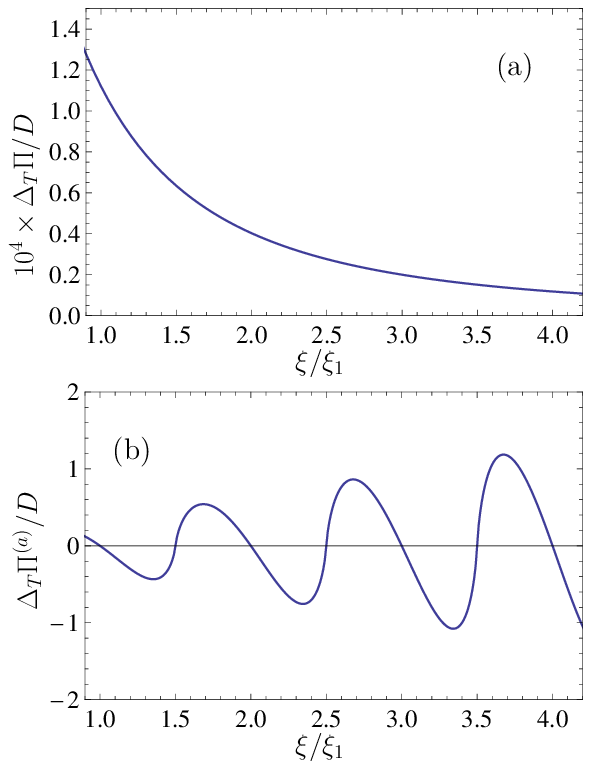}
}
\vspace*{-12cm}
\caption{\label{fg2}(Color online)
The normalized values of the thermal correction (\ref{eq45}) to the combination
of the components
of polarization tensor of graphene computed as functions of imaginary
frequency at $\kb=10\xi_1$ (a) in this paper and (b) using the representation
of Ref.~\cite{10}.
}
\end{figure}
\begin{figure}[b]
\vspace*{-6cm}
\centerline{\hspace*{1cm}
\includegraphics{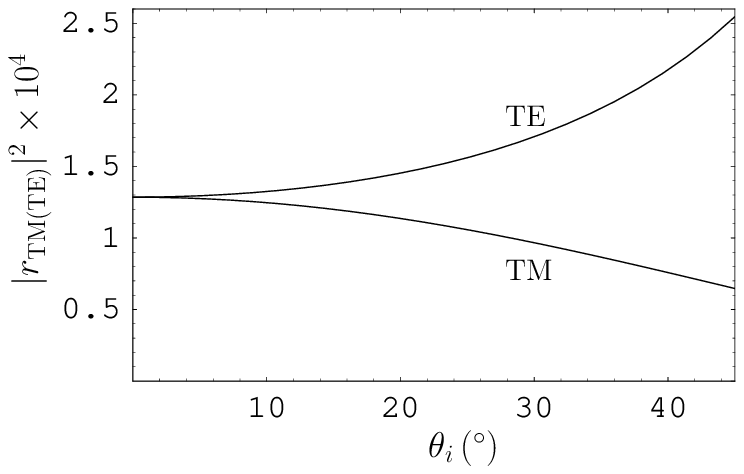}
}
\vspace*{-10cm}
\caption{\label{fg3}
Reflectivities of the TM and TE polarized light of high frequencies on
graphene are shown as functions of the incidence angle by the lower and
upper lines, respectively.
}
\end{figure}
\begin{figure}[b]
\vspace*{-6cm}
\centerline{\hspace*{1cm}
\includegraphics{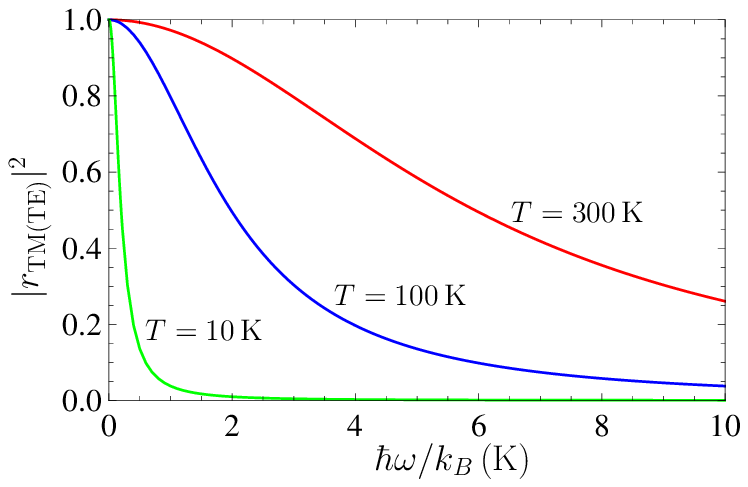}
}
\vspace*{-10cm}
\caption{\label{fg4}(Color online)
Reflectivities of graphene at the normal incidence at different temperatures
$T=10\,$K, 100\,K, and 300\,K are shown as functions of frequency measured in K
by the solid lines from bottom to top,
respectively.
}
\end{figure}
\begin{figure}[b]
\vspace*{-6cm}
\centerline{\hspace*{1cm}
\includegraphics{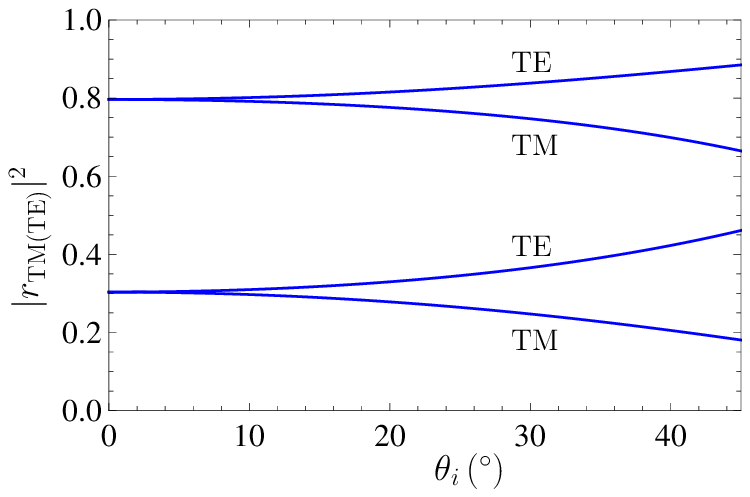}
}
\vspace*{-10cm}
\caption{\label{fg5}(Color online)
Reflectivities of the TM and TE polarized light of low frequencies
($\omega/T=0.01$ and $\omega/T=0.03$ for the upper and lower pairs of
lines, respectively) on
graphene are shown as functions of the incidence angle.
}
\end{figure}
\begin{figure}[b]
\vspace*{-2cm}
\centerline{\hspace*{1cm}
\includegraphics{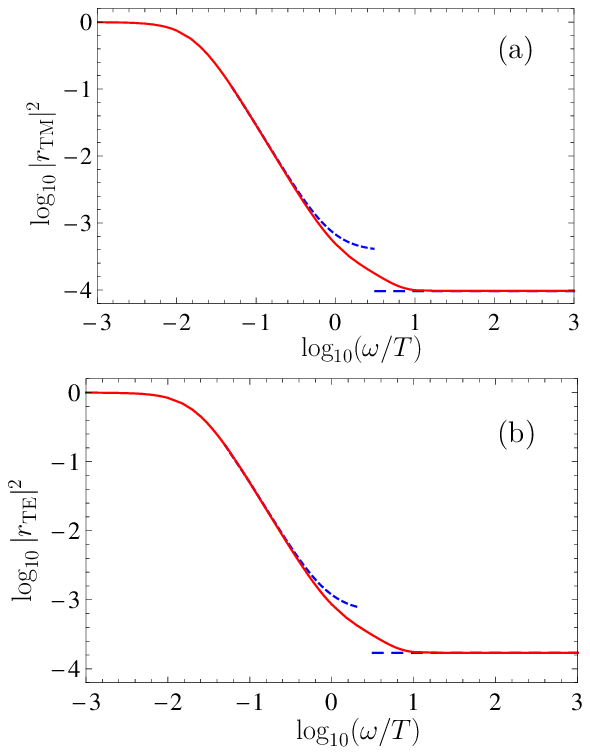}
}
\vspace*{-12cm}
\caption{\label{fg6}(Color online)
Reflectivities of the (a) TM and (b) TE polarized light  on
graphene are shown as functions of $\omega/T$ at the incidence angle
$\theta_i=30^{\circ}$. The solid, short-dashed and long-dashed lines
are computed by the exact formulas, and by the asymptotic expressions
for low and high frequencies, respectively.
}
\end{figure}
\begin{figure}[h]
\vspace*{1cm}
\includegraphics{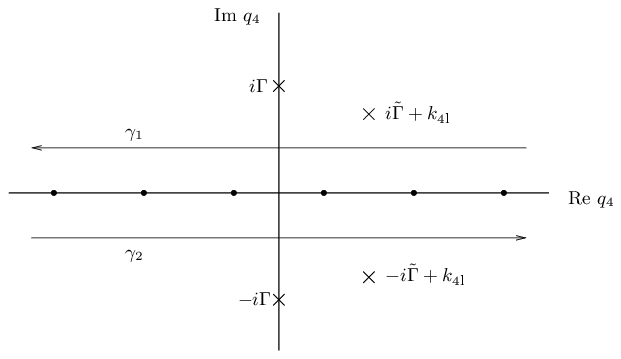}
\vspace*{-16cm}
\caption{\label{last}
The complex $q_4$-plane containing the integration pathes $\gamma_1$ and $\gamma_2$.
The dots indicate the poles at the fermionic Matsubara frequencies and
the crosses indicate four additional poles (see text for further discussion).}
\label{figq4}\end{figure}

\end{document}